\documentclass[12pt,a4paper]{article}
\usepackage{amsmath,bbm,longtable,paralist}
\usepackage[font=scriptsize]{caption}
\usepackage[top=1.15in, bottom=1.15in, left=1.01in, right=1.01in]{geometry}

\DeclareCaptionStyle{italic}[justification=centering]{labelfont={bf},textfont={it},labelsep=colon}
\usepackage{graphicx,psfrag,epsf}
\usepackage{enumerate}
\usepackage{natbib}
\usepackage{url} 
\usepackage{booktabs, subfig, bm, paralist,mathpazo,tikz,todonotes,longtable,microtype,dsfont,rotating,algorithm} 
\usepackage[pdftex,colorlinks=true]{hyperref}
\definecolor{darkblue}{rgb}{0,0,.6}
\hypersetup{citecolor=darkblue,linkcolor=darkblue,urlcolor=darkblue}

\newcommand{\blind}{0}

\newcommand{\X}{\mathcal{X}}
\newcommand{\Y}{\mathcal{Y}}
\graphicspath{{plots/}}

\newsavebox\CBox

\definecolor{a0}{rgb}{0.0, 0.5, 0.0}
\definecolor{bistre}{rgb}{0.24, 0.17, 0.12}
\definecolor{amethyst}{rgb}{0.6, 0.4, 0.8}
\definecolor{blue-violet}{rgb}{0.54, 0.17, 0.89}
\definecolor{Rcolor}{RGB}{150,160,190}
\definecolor{blush}{rgb}{0.87, 0.36, 0.51}
\definecolor{brightturquoise}{rgb}{0.03, 0.91, 0.87}
\definecolor{burntorange}{rgb}{0.8, 0.33, 0.0}

\usepackage{orcidlink}

\date{\today}
\AtBeginDocument{}

\begin{document}

\def\spacingset#1{\renewcommand{\baselinestretch}
{#1}\small\normalsize} \spacingset{1}

\if0\blind
{
  \title{\bf A robust partial least squares approach for function-on-function regression}
  \author{Ufuk Beyaztas \orcidlink{0000-0002-5208-4950} \thanks{Postal address: Department of Statistics, Marmara \"{U}niversitesi G\"{o}ztepe Yerle\c{s}kesi, 34722, Kadik\"{o}y-Istanbul, Turkey; Email: ufuk.beyaztas@marmara.edu.tr}\\
  Department of Statistics\\
   Marmara University \\
  \\
  Han Lin Shang \orcidlink{0000-0003-1769-6430}\\
    Department of Actuarial Studies and Business Analytics \\
    Macquarie University
}
  \maketitle
} \fi

\if1\blind
{
\title{\bf A robust partial least squares approach for function-on-function regression}
} \fi

\bigskip

\begin{abstract}
The function-on-function linear regression model in which the response and predictors consist of random curves has become a general framework to investigate the relationship between the functional response and functional predictors. Existing methods to estimate the model parameters may be sensitive to outlying observations, common in empirical applications. In addition, these methods may be severely affected by such observations, leading to undesirable estimation and prediction results. A robust estimation method, based on iteratively reweighted simple partial least squares, is introduced to improve the prediction accuracy of the function-on-function linear regression model in the presence of outliers. The performance of the proposed method is based on the number of partial least squares components used to estimate the function-on-function linear regression model. Thus, the optimum number of components is determined via a data-driven error criterion. The finite-sample performance of the proposed method is investigated via several Monte Carlo experiments and an empirical data analysis. In addition, a nonparametric bootstrap method is applied to construct pointwise prediction intervals for the response function. The results are compared with some of the existing methods to illustrate the improvement potentially gained by the proposed method.
\\

\end{abstract}

\newpage
\spacingset{1.48}

\section{Introduction} \label{sec:intro}

Over the last decade, the availability of functional data whose sample elements are in the form of curves, images, shapes, or more general manifold-valued objects sampled over a continuum measure has progressively increased in many scientific fields. Consequently, the interest in and need for tools that tackle the analysis of functional datasets is increasing significantly. For example, consult \cite{ramsay2002, ramsay2006}, \cite{ferraty2006}, \cite{Rao2010}, \cite{horvath2012}, \cite{cuevas2014}, \cite{Hsing}, \cite{KoRe}, \cite{Ahmad2020}, and \cite{Israel2020} for recent developments on the theory and applications of functional data analysis tools. 

This paper is devoted to studying the function-on-function linear regression model (FFLRM), where the response and predictors are functional. FFLRM is among the most interesting functional data analysis tools and has received much attention in the statistical literature. This model intends to explore the association between a functional response and one or more functional predictors. In this context, several FFLRMs have been proposed \citep[see, e.g.,][]{ramsay1991, yao2005, ramsay2006, MullerYao2008, matsui2009, he2010, wang2014, ivanescu2015, chiou2016}. In these models, the associations between the functional variables are characterized via the integrated functional predictors weighted by the unknown regression parameter function. Estimating such models is a difficult task since the regression parameter function belongs to an infinite-dimensional space. The common practical approach to overcoming this problem is projecting the regression parameter function onto a finite-dimensional space using a basis expansion technique. For this purpose, several approaches have been proposed. For example, with the general basis expansion methods (i.e., $B$-spline, Fourier, and Gaussian basis function), the traditional estimation methods, such as least-squares (LS) \citep{ramsay1991, ramsay2006}, weighted LS \citep{yamanishi2003}, penalized maximum likelihood \citep{matsui2009}, expectation/conditional maximization either algorithm \citep{wang2014}, and penalized spline \citep{ivanescu2015} have been extended to functional data to estimate the FFLRMs. However, the use of general basis expansion methods may require a large number of basis functions to project the regression parameter function, leading to overfitting of the model. In addition, when a large number of basis functions is used, the aforementioned estimation methods suffer from the multicollinearity problem \citep{PredSap, Agu2016}. Moreover, in such a case, these methods may be computationally intensive \citep{BS19}. Compared with general basis expansion methods, the dimension reduction techniques such as functional principal component (FPC) regression \citep{AguileraPCA, yao2005, Hall2006, chiou2016} and functional partial least squares (FPLS) regression \citep{PredSc, BS19} solve multicollinearity problem and decrease the computation burden in a high-dimensional setting. Thus, the FPC and FPLS are preferable to general basis expansion methods in the estimation of FFLRMs.

In the FPC and FPLS regression methods, the infinite-dimensional functions of the FFLRMs are projected onto the finite-dimensional orthonormal FPC and FPLS bases. In doing so, both methods produce orthogonal latent components. The regression model constructed using the latent components of the functional response and functional predictors are then used to approximate the regression problem of functional response on the functional predictors. The FPC considers maximizing the covariance between the functional predictors to extract the latent components. However, a few FPCs generally contain the most information about the covariance between the functional predictors, which are not necessarily important for representing FPCs. All or some of the most important terms accounting for the interaction between the basis functions and functional predictors might come from later principal components \citep{Delaigle2012}. In contrast, the FPLS components are extracted by maximizing the covariance between the functional response and functional predictors. Thus, compared with FPC, the FPLS produces more informative latent components with fewer terms, which makes FPLS more preferable than FPC \citep{Reiss2007, Delaigle2012, Bande2017}. Therefore in this paper, we consider the FPLS method to estimate the FFLRM.

For the functional regression models, various FPLS methods have been proposed to investigate the relationship between functional variables \citep[see, e.g.,][]{PredSap, HydSh2009, Agu2010, Agu2016, PredSc, BS19, BS21}. In these methods, the FPLS basis functions are iteratively extracted from squared covariance between the functional response and vector of functional predictors to calculate the latent components. The extracted components are then used to estimate the underlying FFLRM. However, the direct estimation of the FPLS basis functions is an ill-posed problem because of the infinite-dimensional structures of the functional random variables. To overcome this problem, \cite{PredSc} and \cite{BS19} proposed approximating the FPLS regression via the finite-dimensional PLS regression of the basis expansion coefficients. The coefficients are obtained by projecting the infinite-dimensional functional random variables onto a finite-dimensional space via a basis expansion method. \cite{PredSc} and \cite{BS21} showed that the same PLS components are obtained under both the FPLS and the PLS regression constructed using the basis expansion coefficients. The numerical results provided by the studies above showed that the FPLS produces an improved predictive performance over the existing methods because the extracted PLS components consider the correlation between the response and predictor variables.

Most of the existing methods, including FPLS, used to estimate the FFLRMs are non-robust to outlying observations, which are generated by a stochastic process with a distribution different from that of the vast majority of the remaining observations \citep[see, e.g.,][]{Rana}. In functional data, there are three types of outliers:
\begin{inparaenum}
\item[(1)] \textit{magnitude outliers}, which are points far from the bulk of the data; 
\item[(2)] \textit{shape outliers}, which fall within the range of the data but differ in shape from the bulk of the data; and 
\item[(3)] the combination of magnitude and shape outliers \citep[e.g., see][]{Beyaztas}. 
\end{inparaenum}
In this study, we consider the combination of magnitude and shape outliers. In the case of outliers, the LS or maximum likelihood-based estimation methods produce biased estimates; thus, predictions obtained from the fitted model become unreliable \citep[see, e.g.,][]{Jun2018}. To overcome this issue, several robust methods have been proposed in functional regression models. In the scalar-on-function regression model (where the response is a scalar random variable while the predictors consist of random curves), see for example, \cite{Maronna}, \cite{ShinLee}, and \cite{Kalogridis}. In the context of FFLRM, \cite{Gervini} proposed an outlier-resistant loss function-based robust method. Also, \cite{Harjit} proposed a Fisher-consistent robust regression model, and they proposed a robust functional principal component (RFPC)-based estimator to estimate the model parameter. However, these two methods are based on the simple FFLRM, with only one predictor in the model. To the best of our knowledge, there is no method to robustly estimate the FFLRM by extracting information from both the response and multiple predictors.

This paper proposes a robust FPLS method, which allows for more than one functional predictor in the model to estimate the FFLRM. Similar to the FPLS methods of \cite{PredSc} and \cite{BS19}, in the proposed method, the infinite-dimensional FPLS regression model is approximated via the finite-dimensional PLS regression of the basis expansion coefficients. In the finite-dimensional case, the PLS components are obtained via several algorithms, such as nonlinear iterative PLS \citep{nipals}, SIMPLS \citep{simpls}, and improved kernel PLS \citep{dayal}. Our proposed method restricts our attention to SIMPLS, which deflates the empirical covariance matrix to obtain the PLS components. The traditional SIMPLS algorithm uses an LS-type estimator to extract the PLS components. Thus, it may seriously be affected by outlying observations and lead to unreliable estimation and prediction results. Several robust PLS algorithms have been developed to deal with outliers. For example, \cite{Wakelinc} and \cite{Cummins} proposed robust PLS algorithms using robust iteratively reweighted algorithms. However, these methods use non-robust initial weights and are not robust to outliers in the predictor space (called leverage points). Using a robust covariance matrix with the SIMPLS algorithm, \cite{Hubert} proposed a robust PLS regression. \cite{Serrneels} developed a robust PLS algorithm based on the M-regression estimator. In addition, using the weight function developed by \cite{Markatou1996}, \cite{AA17} proposed a robust iteratively reweighted SIMPLS method (IRSIMPLS). The latter methods are robust to outliers in the predictor space (vertical outliers) and leverage points. The numerical analyses performed by \cite{AA17} demonstrate that the IRSIMPLS produces improved results compared with its competitors. Therefore, we consider the IRSIMPLS algorithm in the proposed method to robustly approximate the infinite-dimensional FPLS regression model in the presence of outliers. In the numerical analyses given in Sections~\ref{sec:mce}, the predictive performance of the proposed method is examined via several Monte Carlo experiments, and the results are compared with those of existing methods. Our numerical results show that the proposed method is robust to outliers and improves prediction performance compared to the available methods when outliers are present in the data. Also, our results show that the proposed robust FPLS approach produces competitive or even better results with those of the RFPC method of \cite{Harjit}.

The remainder of this paper is organized as follows. An overview of the FPLS approach and the methodology of the proposed method are presented in Section~\ref{sec:fpls}. A series of Monte Carlo experiments are conducted to evaluate the finite-sample performance of the proposed method, and the results are given in Section~\ref{sec:mce}. The proposed method is used to analyze an empirical data analysis and the results presented in Section~\ref{sec:eda}. Section~\ref{sec:conc} concludes the paper, along with some ideas on how the methodology can be further extended.

\section{The functional PLS} \label{sec:fpls}

Let $\Y(t)$ and $\X_m(s)$, for $m = 1, \ldots, M$, respectively, denote a response and $M$ sets of predictors given as functions with closed and bounded intervals $t \in T$ and $s \in S$ as domains. We assume that both the functional response and the functional predictors are second-order stochastic processes with finite second-order moments, and are the elements of the $\mathcal{L}_2$ separable Hilbert space, the space of square-integrable and real-valued functions. Without loss of generality, we also assume that $\Y(t)$ and $\X_m(s)$ for $m = 1, \ldots, M$, are zero-mean stochastic processes so that $\mathbb{E} \left[ \Y(t) \right] = \mathbb{E} \left[ \X_m(s) \right] = 0$. 

We consider the following FFLRM for exploring the association between the functional response and $M$ sets of functional predictors:
\begin{equation}\label{eq:fof_ed}
\Y(t) = \int_S \pmb{\X}^\top(s) \pmb{\beta}(s,t) \textrm{d}s + \epsilon(t),
\end{equation}
where $\pmb{\X}(s) = \left[ \X_1(s), \ldots, \X_M(s) \right]^\top$ is the matrix-valued functions comprising of $M$ sets of functional predictors, $\pmb{\beta}(s,t) = \left[ \beta_1(s,t), \ldots, \beta_M(s,t) \right]^\top$ denotes the vector of bivariate coefficient functions, and $\epsilon(t)$ is the error function which is assumed to be independent of $\X_m(t)$s. The primary aim in~\eqref{eq:fof_ed} is to estimate the regression parameter function $\pmb{\beta}(s,t)$.

The FPLS components associated with the FFLRM in~\eqref{eq:fof_ed} can be obtained as the solutions of Tucker's criterion \citep{Tucker} extended to functional variables as follows:
\begin{equation}\label{eq:tucker}
\underset{\begin{subarray}{c}
  \kappa \in \mathcal{L}_2,~ \Vert \kappa_m \Vert_{\mathcal{L}_2} = 1, \forall m \in [1, \ldots, M] \\
  \zeta \in \mathcal{L}_2,~ \Vert \zeta \Vert_{\mathcal{L}_2} = 1
  \end{subarray}}{\max} \text{Cov}^2 \left( \int_S \pmb{\X}^\top(s) \bm{\kappa}(s) ds, ~ \int_T \Y(t) \zeta(t) dt \right),
\end{equation}
where $\bm{\kappa}(s) = [\kappa_1(s), \ldots, \kappa_m(s)]^\top$ and $\zeta(t)$ denote weight functions associated with the predictor and response variables, respectively.

As noted by \cite{PredSap}, the FPLS components are also the eigenvectors associated with the largest eigenvalues of the product of Escoufier operators \citep{Escoufier}. The first FPLS component of the FFLRM of $\Y(t)$ on $\pmb{\X}(s)$, denoted by $\bm{\eta}_1$, is obtained by conducting the ordinary linear regressions of $\pmb{\X}(s)$ on $\bm{\eta}_1$ and $\Y(t)$ on $\bm{\eta}_1$ as follows:
\begin{equation*}
\bm{\eta}_1 = \int_S \pmb{\X}^\top(s) \bm{\kappa}_1(s) \textrm{d}s,
\end{equation*}
where $\bm{\kappa}_1(s)$, the weight function associated with the first FPLS component, is defined as follows:
\begin{equation*}
\bm{\kappa}_1(s) = \frac{\int_T \mathbb{E} \left[ \pmb{\X}(s) \Y(t) \right] \textrm{d}t}{\sqrt{\int_S \left( \int_T \mathbb{E} \left[ \pmb{\X}(s) \Y(t) \right] \textrm{d}t \right)^2 \textrm{d}s}}.
\end{equation*}

FPLS is an iterative method, and thus, the subsequent FPLS components are computed via an iterative stepwise procedure. Let $h = 1, 2, \ldots$ denote the iteration number. In addition, let $\pmb{\X}_0(s) = \pmb{\X}(s)$ and $\Y_0(t) = \Y(t)$. Denote by $\pmb{\X}_h(s)$ and $\Y_h(t)$ the residuals of the following regression models:
\begin{align*}
\pmb{\X}_h(s) &= \pmb{\X}_{h-1}(s) - \bm{p}_h(s) \bm{\eta}_h, \\
\Y_h(t) &= \Y_{h-1}(t) - \zeta_h(t) \bm{\eta}_h,
\end{align*}
where $\bm{p}_h(s) = \frac{\mathbb{E} \left[ \pmb{\X}_{h-1}(s) \bm{\eta}_h\right]}{\mathbb{E} \left[ \bm{\eta}_h^2 \right]}$, and $\zeta_h(t) = \frac{\mathbb{E} \left[ \Y_{h-1}(t) \bm{\eta}_h \right]}{\mathbb{E} \left[ \bm{\eta}_h^2 \right]}$. Then, the $h$\textsuperscript{th} FPLS component, denoted by $\bm{\eta}_h$, is obtained by maximizing the Tucker's criterion~\eqref{eq:tucker}:
\begin{equation*}
\bm{\eta}_h = \int_S \pmb{\X}_{h-1}^\top(s) \bm{\kappa}_h(s) \textrm{d}s,
\end{equation*}
where the weight function $\bm{\eta}_h$ is given by:
\begin{equation*}
\bm{\kappa}_h(s) = \frac{\int_T \mathbb{E} \left[ \pmb{\X}_{h-1}(s) \Y_{h-1}(t) \right] \textrm{d}t}{\sqrt{\int_S \left( \int_T \mathbb{E} \left[ \pmb{\X}_{h-1}(s) \Y_{h-1}(t) \right] \textrm{d}t \right)^2 \textrm{d}s}}.
\end{equation*}
At the final step, the FPLS regression is completed by conducting the linear regressions of $\pmb{\X}_{h-1}(s)$ and $\Y_{h-1}(t)$ on $\bm{\eta}_h$.

Practically, the FPLS components of the FFLRM in~\eqref{eq:fof_ed} are obtained based on the randomly observed sample curves of the functional response and functional predictor variables. Although these sample curves are intrinsically infinite-dimensional, they are observed in a set of discrete-time points in practice. For this reason, obtaining the sample estimates of the Escoufier's operators becomes problematic because they need to be estimated from the discretely observed observations \citep[see][]{Agu2010}. To address this issue, \cite{PredSc} and \cite{BS19} proposed to use standard PLS regression of the basis expansion coefficients of the functional variables to approximate the FPLS components. In this context, several basis expansion methods, such as $B$-spline, Fourier, wavelet, radial, and Bernstein polynomial bases, have been proposed to project infinite-dimensional functional variables onto the finite-dimensional space. In this study, we consider the $B$-spline basis expansion coefficients of the functional variables, which are then modeled using the IRSIMPLS regression of \cite{AA17} to approximate the FPLS regression of $\Y(t)$ on $\pmb{\X}(s)$.

The IRSIMPLS is a weighted counterpart of the usual SIMPLS \citep[see, e.g.,][for more information about the SIMPLS and IRSIMPLS algorithms]{AA17}, where the weights are calculated based on the weighted likelihood methodology of \cite{Markatou1996}. The weighted likelihood uses weighted score equations to measure the discrepancy between the estimated and hypothesized model densities.

Let us consider the FFLRM in~\eqref{eq:fof_ed} and let $K_{\Y}$ and $K_{\pmb{\X}}$ denote the numbers of basis functions used for approximating the response and predictor functions, respectively. Then, they can be represented as basis function expansion as follows:
\begin{align*}
\Y(t) &= \sum_{k=1}^{K_{\Y}} c_k \phi_k(t) = \pmb{C}^\top \pmb{\Phi}(t), \\
\pmb{\X}(s) &= \sum_{j=1}^{K_{\pmb{\X}}} \bm{d}_j \bm{\psi}_j(s) = \pmb{D}^\top \pmb{\Psi}(s),
\end{align*}
where $\pmb{\Phi}(t) = \left[\phi_1(t), \ldots, \phi_{K_{\Y}}(t) \right]^\top$ and $\pmb{\Psi}(s) = \left[ \bm{\psi}_1(s), \ldots, \bm{\psi}_{M}(s) \right]^\top$ with $\bm{\psi}_m(s) = \left[\psi_{m1}(s), \ldots, \psi_{mK_{\pmb{\X}}}(s) \right]^\top$ respectively denote the basis functions, and $\pmb{C} = \left[ c_1, \ldots, c_{K_{\Y}} \right]^\top$ and $\pmb{D} = \left[ \bm{d}_1, \ldots, \bm{d}_{M} \right]^\top$ with $\bm{d}_m = \left[ d_{m1}, \ldots, d_{mK_{\pmb{\X}}} \right]^\top$ are the corresponding coefficient matrices.

Let $\pmb{\Phi} = \int_T \pmb{\Phi}(t) \pmb{\Phi}^\top(t) dt$ and $\pmb{\Psi} = \int_S \pmb{\Psi}(s) \pmb{\Psi}^\top(s) ds$, respectively, denote the $K_{\Y} \times K_{\Y}$ and $K_{\pmb{\X}} \times K_{\pmb{\X}}$ dimensional symmetric block-diagonal matrices of the inner products of the basis functions. Denote by $\pmb{\Phi}^{1/2}$ and $\pmb{\Psi}^{1/2}$ the square roots of $\pmb{\Phi}$ and $\pmb{\Psi}$, respectively. Then, we consider the following regression model to approximate the FPLS regression of $\Y(t)$ on $\pmb{\X}(s)$:
\begin{equation}\label{eq:model_pls}
\pmb{\Lambda} = \pmb{\Pi} \pmb{\Omega} + \pmb{\epsilon},
\end{equation}
where $\pmb{\Lambda} = \pmb{C}^\top \pmb{\Phi}^{1/2}$, $\pmb{\Pi} = \pmb{D}^\top \pmb{\Psi}^{1/2}$, and $\pmb{\Omega}$ and $\pmb{\epsilon}$ denote the coefficient and residual matrices, respectively. In~\eqref{eq:model_pls}, the between columns of the obtained design matrices ($\pmb{\Lambda}$ and $\pmb{\Pi}$) become highly correlated (multicollinearity) due to the nature of functional data. In addition, even though the functional predictors are independent, because of the high-dimensional structures of the design matrices, there may be a strong correlation between $\pmb{\Lambda}$ and $\pmb{\Pi}$. The traditional estimation techniques such as LS and maximum likelihood extended to functional data suffer from the multicollinearity problem. They may not produce a valid estimate for the coefficient matrix $\pmb{\Omega}$ \citep[see, e.g.,][]{BS19}. On the other hand, the PLS approach can produce an efficient estimate for $\pmb{\Omega}$ by eliminating the multicollinearity problems. It has been showed by \cite{PredSc} and \cite{BS21} that the FPLS regression of $\Y(t)$ on $\pmb{\X}(s)$ is equivalent to the PLS regression of $\pmb{\Lambda}$ on $\pmb{\Pi}$ in the sense that at each step $h$ of the PLS algorithm $1 \leq h \leq H$, the same PLS components are obtained for both PLS regressions. $H$ denotes the retained number of PLS components. 

We consider the IRSIMPLS of \cite{AA17} to approximate the coefficient function $\pmb{\beta}(s,t)$ in~\eqref{eq:fof_ed}. Similar to the traditional PLS algorithms, the IRSIMPLS assumes a linear correlation between predictors and the response and bypasses the multicollinearity problem. In the IRSIMPLS, we assume that the theoretical residuals $(\pmb{\epsilon}_1,\dots,\pmb{\epsilon}_n)$ belong to a parametric family of densities $\mathcal{F} = \left\lbrace f(\cdot, \sigma); \sigma \in \mathbb{R}^+ \right\rbrace$ where $f(\cdot, \sigma)$ is a random variable with mean zero corresponding to given $\sigma$. Let $\widehat{\pmb{\epsilon}}_i$ denote the calculated residuals for the given estimates $\widehat{\pmb{\Omega}}$ and $\widehat{\sigma}$, appropriately determined by the density $f(\cdot, \widehat{\sigma})$. Denote by $g^*(\widehat{\pmb{\epsilon}}_i)$ and $f^*(\widehat{\pmb{\epsilon}}_i; \sigma)$ the kernel density estimator and the smoothed model density, respectively, as follows:
\begin{align*}
g^*(\widehat{\pmb{\epsilon}}_i) &= \int \tau \left( \widehat{\pmb{\epsilon}}_i; u, v \right) \textrm{d} \widehat{F}(u), \\
f^*(\widehat{\pmb{\epsilon}}_i; \sigma) &= \int \tau \left( \widehat{\pmb{\epsilon}}_i; u, v \right) \textrm{d} \mathcal{F}(u; \sigma),
\end{align*}
where $\widehat{F}(u)$ denotes the empirical distribution function of the calculated residuals $\widehat{\pmb{\epsilon}}_i$ for $i = 1, \ldots, n$ and $\tau \left( \widehat{\pmb{\epsilon}}_i; u, v \right)$ is the Gaussian kernel density as follows:
\begin{equation*}
\tau \left( \widehat{\pmb{\epsilon}}_i; u, v \right) = \frac{e^{-(\widehat{\pmb{\epsilon}}_i - u)^2 / (2 v^2)}}{\sqrt{2 \pi} v}. 
\end{equation*}
Herein, $v$ denotes the smoothing parameter, and in this study, it is chosen as $v = \sqrt{\gamma \sigma^2}$, where $\gamma$ is a constant, to assign small weights to outliers \citep{Markatou1996, AA17}. Then, the Pearson residuals $\delta \left( \widehat{\pmb{\epsilon}}_i \right)$ for $i = 1, \ldots, n$, which are used to calculate weights, are defined as follows:
\begin{equation*}
\delta \left( \widehat{\pmb{\epsilon}}_i \right) = \frac{g^*(\widehat{\pmb{\epsilon}}_i)}{f^*(\widehat{\pmb{\epsilon}}_i; \sigma)}.
\end{equation*}
Finally, the weight for $i$\textsuperscript{th} observation, $\omega_i$, is defined as follows:
\begin{align*}
\omega_i &= \omega \left( \widehat{\pmb{\epsilon}}_i; f^*(\widehat{\pmb{\epsilon}}_i; \sigma), \widehat{F} \right)  \\
&= \min \left\lbrace 1, \frac{\left[ A \left( \delta \left(\widehat{\pmb{\epsilon}}_i\right) \right) + 1\right]^+ }{\delta \left(\widehat{\pmb{\epsilon}}_i\right) + 1} \right\rbrace,
\end{align*}
where $A(\cdot)$ denotes the residual adjustment function. In this study, the weights are calculated based on the Hellinger residual adjustment function of \cite{Lindsay}: $A[\delta(\cdot)] = 2 [\delta(\cdot) + 1]^{1/2} - 1$.

The IRSIMPLS produces PLS components by deflating a weighted covariance matrix $\pmb{S}^{\omega} = \left( \pmb{\Pi}^{\omega} \right)^\top \pmb{\Lambda}^{\omega}$. The $\pmb{\Pi}^{\omega}$ and $\pmb{\Lambda}^{\omega}$ are the weighted versions of $\pmb{\Pi}$ and $\pmb{\Lambda}$, respectively, and obtained by multiplying each row of $\pmb{\Pi}$ and $\pmb{\Lambda}$ by the square root of weights:
\begin{equation*}
\omega_i = \underset{k}{\text{median}} \lbrace \omega_{ik} \rbrace, \qquad i = 1, \ldots, n,\quad k = 1, \ldots, K_{\Y}.
\end{equation*}
Herein, $ \omega_{ik}$ denotes the the weight for $i$\textsuperscript{th} residual corresponding to the $k$\textsuperscript{th} column. The algorithm of the IRSIMPLS is described in Algorithm 1.

\begin{algorithm}
\begin{itemize}
\item[Step 1.] Center the matrices $\pmb{\Pi}$ and $\pmb{\Lambda}$ using $L_1$ median as defined by \cite{Serrneels}. In addition, scale the centered $\pmb{\Pi}$ using median absolute deviation. Let $\pmb{\Pi}^*$ denote the centered and scaled version of $\pmb{\Pi}$.
\item[Step 2.] For $i = 1, \ldots, n$, compute weights $\omega_i$ for each row of $\pmb{\Pi}^*$, and obtain $\pmb{\Pi}^{\omega}$ and $\pmb{\Lambda}^{\omega}$ by multiplying each row of $\pmb{\Pi}$ and $\pmb{\Lambda}$ by the square roots of $\omega_i$.
\item[Step 3.] Perform ordinary SIMPLS on a sample of $n_r$ ($n_r \ll n$) randomly drawn pairs without replacement from $\left( \pmb{\Pi}, \pmb{\Lambda}\right)$ to obtain starting coefficient estimate $\widehat{\pmb{\Omega}}^*$. Note that the starting values to calculate $\widehat{\pmb{\Omega}}^*$ are determined by applying ordinary SIMPLS on the centered and scaled $\pmb{\Pi}$ and $\pmb{\Lambda}$.
\item[Step 4.] Compute residuals using the starting coefficient estimate obtained in the previous step as follows:
\begin{equation*}
\widehat{\pmb{\epsilon}}_i = \pmb{\Lambda} - \pmb{\Pi} \widehat{\pmb{\Omega}}^*.
\end{equation*}
\item[Step 5.] Obtain the centered and scaled residuals $\widehat{\pmb{\epsilon}}^*_i$ using $L_1$ median and median absolute deviation.
\item[Step 6.] Calculate new weights, $\omega^*_i$, using the residuals $\widehat{\pmb{\epsilon}}^*_i$ obtained in Step 5 to reweight $\pmb{\Pi}$ and $\pmb{\Lambda}$ multiplying each row of them by the square roots of $\omega^*_i$. Let $\pmb{\Pi}^{\omega*}$ and $\pmb{\Lambda}^{\omega*}$ denote the reweighted versions of $\pmb{\Pi}$ and $\pmb{\Lambda}$, respectively.
\item[Step 7.] Obtain the reweighted covariance matrix $\pmb{S}^{\omega*} = \left( \pmb{\Pi}^{\omega*} \right)^\top \pmb{\Lambda}^{\omega*}$. Then, compute the new coefficient estimate $\widehat{\pmb{\Omega}}^{**}$ performing the ordinary SIMPLS on $\pmb{S}^{\omega*}$.
\item[Step 8.] With the coefficient estimate in Step 7, repeat steps 4-7 until convergence is achieved. Convergence is reached if the maximum of the absolute deviations between two $\widehat{\pmb{\Omega}}^{**}$s obtained consecutive repeats less than a predetermined threshold, such as $10^{-4}$.
\item[Step 9.] Repeat Steps 3-8 a few times to obtain the final estimate of $\pmb{\Omega}$, given by $\pmb{\Omega}^{\omega}$, which is equal to one having the minimum absolute deviation among others.
\end{itemize}
\caption{IRSIMPLS algorithm}
\label{alg:pls}
\end{algorithm}

Let $\pmb{\Omega}^{\omega,h}$ denote the estimated coefficient matrix of $\pmb{\Omega}$, obtained after $h$ iterations from the IRSIMPLS regression. Then, the IRSIMPLS approximation of $\pmb{\beta}(s,t)$, given by $\widehat{\pmb{\beta}}^{\omega}(s,t)$, is obtained using the bases $\pmb{\Phi}(t)$ and $\pmb{\Psi}(s)$ as follows:
\begin{equation}\label{eq:irw_approx}
\widehat{\pmb{\beta}}^{\omega}(s,t) = \left[ \left( \pmb{\Psi}^{1/2} \right)^{-1} \pmb{\Omega}^{\omega,h} \left( \pmb{\Phi}^{1/2} \right)^{-1} \right] \pmb{\Psi}(s) \pmb{\Phi}(t).
\end{equation}
In what follows, the robust estimation of the functional response, $\widehat{\Y}(t)$, can be calculated as follows:
\begin{equation*}
\widehat{\Y}(t) = \int_S \bm{\X}(s) \widehat{\pmb{\beta}}^{\omega}(s,t) \textrm{d}s.
\end{equation*}
Since the IRSIMPLS approximation of $\pmb{\beta}(s,t)$ given in~\eqref{eq:irw_approx} is based on the weighted residuals, it is more robust to the outliers than the ordinary LS-based PLS methods.

The performance of the proposed method depends on the number of PLS components $h$ used to estimate Model~\eqref{eq:fof_ed}. Thus, the optimum number $h$ should be determined based on an error metric. The outliers may affect the usual error metrics, leading to incorrectly determining the optimum number $h$. Therefore, we consider a trimmed mean absolute prediction percentage error (TMAPE) to determine the optimum $h$ robustly. To this end, we consider the following procedure. First, the entire dataset is randomly divided into two parts roughly the same size. The first part of the data is used to construct the FFLRM, and the PLS approximation of $\pmb{\beta}(s,t)$ is obtained for $h = 1, 2, \ldots, H$ number of PLS components. With the second part of the data, the absolute prediction percentage errors (APE) are calculated as follows:
\begin{equation*}
\text{APE}_i(h) = \left\Vert \frac{\left\vert \widehat{\Y}_i(t) - \Y_{i}(t) \right\vert}{\left\vert \Y_{i}(t) \right\vert} \right\Vert_{\mathcal{L}_2} \qquad i = 1, 2, \ldots, \lfloor n/2 \rfloor,
\end{equation*}
where $\left\Vert \cdot \right\Vert_{\mathcal{L}_2}$ denotes $\mathcal{L}_2$ norm, which is approximated by the Riemann sum \citep{LuoQi}, $\lfloor \cdot \rfloor$ denotes an integer value that rounds down to $n/2$, and $\Y_{i}(t)$ and $\widehat{\Y}_i(t)$ are the observed response functions for $i^{\text{th}}$ individual and its prediction, respectively. Then, the following TMAPE is computed for each $h$:
\begin{equation*}
\text{TMAPE}(h) = \frac{1}{n*} \sum_{i=1}^{n*} \text{APE}_i(h),
\end{equation*}
where $\left\lbrace n* \subset \lbrace 1, 2, \ldots, \lfloor n/2 \rfloor \rbrace,~ \vert n* \vert = \left[q  \lfloor n/2 \rfloor \right] \right\rbrace$. Herein, we consider 20\% trimming proportion as suggested by \cite{Wilcox} so that $q = 0.8$ (other trimming proportions such as 5\% and 10\% can also be used in the proposed method). In this way, the optimum $h$ is robustly determined by ignoring the APE values corresponding to the outliers. Finally, the optimum number of PLS components is determined according to the minimum TMAPE.

\section{Monte Carlo experiments} \label{sec:mce}

Computationally, the proposed method works with a similar structure with the FPLS methods of \cite{PredSc} and \cite{BS19}. We note that all the numerical calculations in this study are performed using \texttt{R~4.1.1.} on an Intel Core i7 6700HQ 2.6 GHz PC. We summarize the steps and \texttt{R} packages used in the computation of the proposed method as follows. In the first step of the proposed method, the $B$-spline basis expansion method is used to project the functional random variables onto the finite-dimensional space, which can easily be done using \texttt{create.bspline.basis} function in the \texttt{R} package \texttt{fda} \citep{fda}. In addition, the \texttt{inprod} and \texttt{sqrtm} functions in the \texttt{R} packages \texttt{fda} and \texttt{expm} \citep{expm} are used to obtain the inner product matrices and their square roots, respectively. In the second step, the IRSIMPLS algorithm proposed by \citep{AA17} uses the \texttt{R} package \texttt{wle} \citep{wlep} to obtain the weights. These weights are used to robustly estimate the regression parameter matrix of the PLS regression constructed using basis expansion coefficients. The estimated parameter matrix and $B$-spline basis functions are then used to obtain the approximate form of the regression parameter function $\bm{\beta}(s,t)$.

Various Monte Carlo experiments under different data generating processes are conducted to evaluate the finite-sample performance of the proposed method. The results are compared with those obtained via the LS, na\"ive SIMPLS, and the RFPC of \cite{Harjit}. Throughout the experiments, the following FFLRM is considered:
\begin{equation}\label{eq:model_MC}
\Y(t) = \sum_{m=1}^{M=5} \int_S \X_m(s) \beta_m(s,t) \textrm{d}s + \epsilon(t),
\end{equation}
where $s, t \in [0, 1]$. Two cases are considered for generating the data:
\begin{description}
\item[Case-1:] The functions of both the response and predictor variables are generated from smooth processes with no outliers. In this case, the proposed method is expected to produce similar results with available traditional methods. Our aim, in this case, is to show the correctness of the proposed method.
\item[Case-2:] The functional variables of the response and predictor are contaminated by inserting outliers at 20\% contamination level. In this case, the proposed method is expected to produce better prediction performance than the traditional methods.
\end{description}

We consider two scenarios to generate the functions of the predictor variables. In the first scenario (Scenario-1), the functional predictors are assumed to be independent (i.e., $\text{Cov}(\X_m(s), \X_{m^{\prime}}(s)) = 0$ for $m \neq m^{\prime}$ and $m, m^{\prime} = 1, \ldots, 5$) and are generated as follows:
\begin{equation*}
\X_m(s) = 10 + V_m(s),
\end{equation*}
where $V_m(s)$ are generated from the Gaussian process with mean zero and variance-covariance function $\pmb{\Sigma}_V(s, s^{\prime}) = e^{-100 (s - s^{\prime})^2}$. In the second scenario (Scenario-2), on the other hand, the functional predictors are generated as follows:
\begin{equation*}
\X_m(s) = 10 + \sum_{j = 1}^{\text{Lag}} V_{m+j}(s) / \sqrt{\text{Lag}+1},
\end{equation*}
where $V_m(s)$ ($m = 1, \ldots, 9$) are generated as in the first scenario and Lag controls the correlation between the functional predictors \citep{LuoQi}. In the experiments, similar to \cite{LuoQi}, we consider Lag = 4 to generate correlated functional predictors. The functions of the response variable are obtained using~\eqref{eq:model_MC} with the following coefficient functions:
\begin{align*}
\beta_1(s,t) &= (1-s)^2 (t-0.5)^2,\\
\beta_2(s,t) &= e^{-3 (s-1)^2} e^{-5 (t-0.5)^2},\\
\beta_3(s,t) &= e^{-5(s-0.5)^2 -5(t-0.5)^2} + 8e^{-5(s-1.5)^2 -5(t-0.5)^2},\\
\beta_4(s,t) &= \sin(1.5 \pi s) \sin(\pi t),\\
\beta_5(s,t) &= \sqrt{s t}.
\end{align*}

\begin{figure}[!htb]
  \centering
  \includegraphics[width=8.6cm]{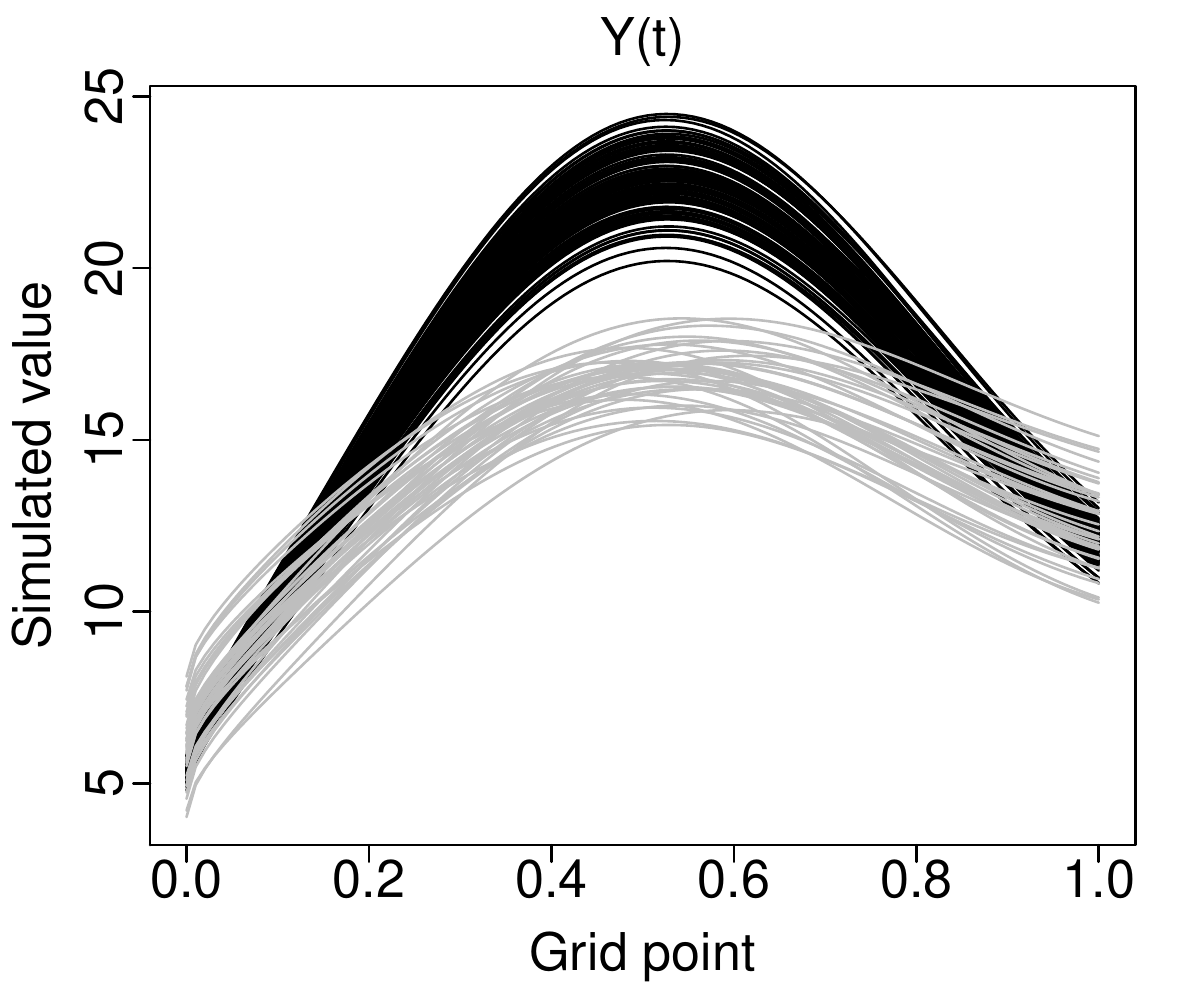}
\quad
  \includegraphics[width=8.6cm]{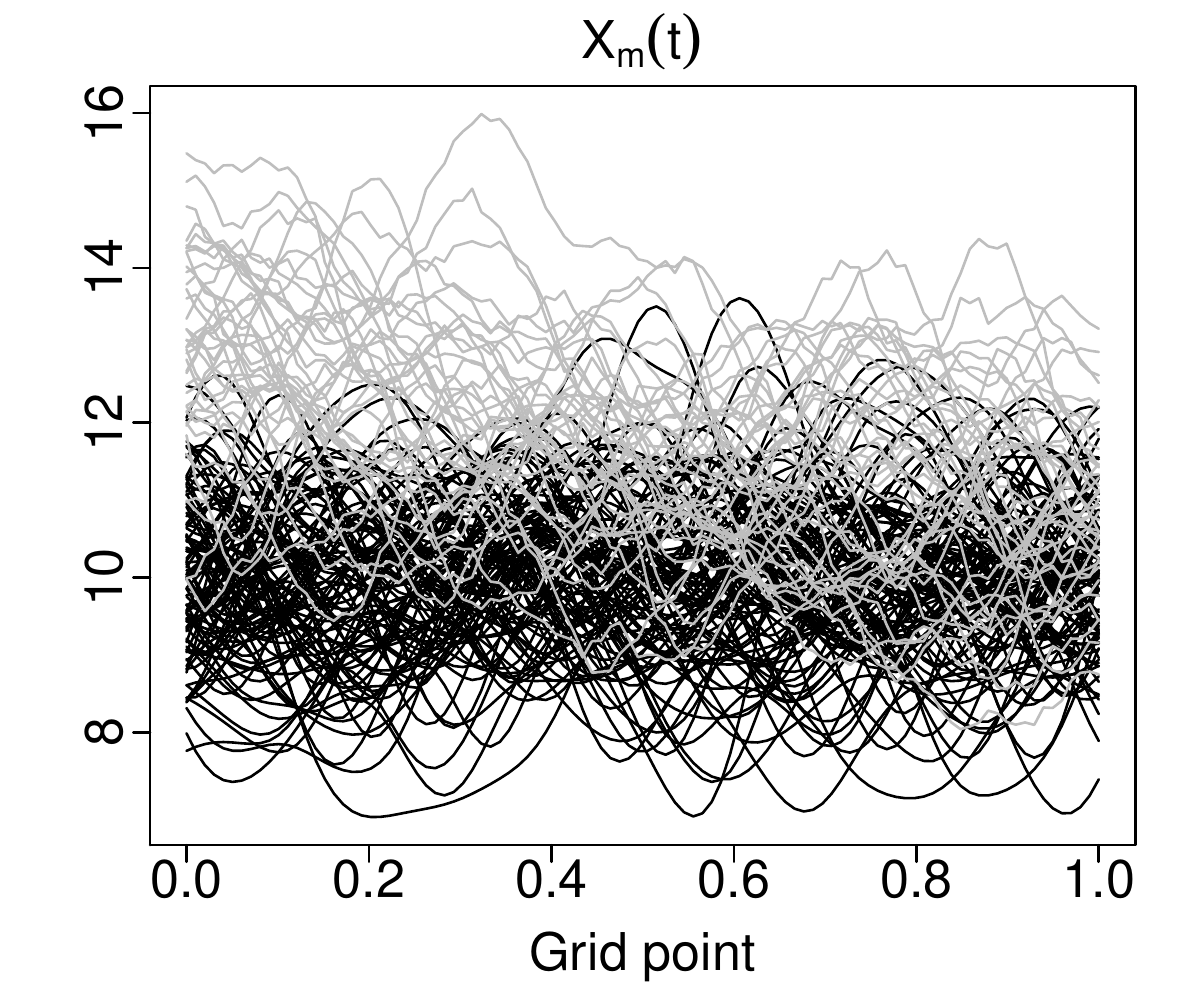}
  \caption{\small{Plots of the generated functions of the response variable (left panel) and predictor variables (right panel). Curves in black color present the non-outlying functions, while gray curves present the generated outlying functions}}\label{fig:Fig_1}
\end{figure}

For each variable, $n = 500$ functions are generated at 100 equally spaced points in the interval $[0,1]$. Then, to evaluate the out-of-sample predictive performance of the methods, the generated data are divided into the following training and test samples with sample sizes $n_{\text{train}}$ and $n_{\text{test}}$, respectively:
\begin{inparaenum}
\item[1)] the first $n_{\text{train}} = 200$ functions are used to construct the FFLRM, and
\item[2)] the last $n_{\text{test}} = 300$ functions are used to predict the response based on the constructed model and a new set of predictors.
\end{inparaenum}
To generate the outlier-contaminated data, $\left( 0.2 \times n_{\text{train}} \right)$ randomly selected functions of the predictors are replaced by $\widetilde{\X}_m(s) = \X_m(s) + \varepsilon_m(s)$, where $\varepsilon_m(s)$ is generated from the Ornstein-Uhlenbeck process:
\begin{equation*}
\varepsilon_m(s) = \gamma + [\varepsilon_{m0}(s) - \rho] e^{-\theta s} \sigma \int_0^s e^{-\theta (s - u)} \textrm{d}W_u,
\end{equation*}
where $\rho$, $\theta$, and $\sigma > 0$ are constants, and $W_u$ is the Wiener process. Note that the initial values $\varepsilon_{m0}(s)$ are independently taken from $W_u$. Then, the outlying functions of the response variable are obtained using~\eqref{eq:model_MC} but with $\widetilde{\X}_m(s)$ and the following modified regression coefficient functions (only $\beta_2(s,t)$, $\beta_3(s,t)$, and $\beta_4(s,t)$ are modified):
\begin{align*}
\widetilde{\beta}_2(s,t) &= e^{-3 (s-1)^2} e^{-3 (t-0.5)^2},\\
\widetilde{\beta}_3(s,t) &= 6 e^{-5(s+0.5)^2 -5(t-0.5)^2} + 4e^{-5(s-1.5)^2 -5(t-0.5)^2},\\
\widetilde{\beta}_4(s,t) &= 6 \cos(6 \pi s) \cos(\pi t).
\end{align*} 
In this way, the functional predictors are contaminated by the magnitude outliers; the response variable is contaminated by the mixture of the magnitude and shape outliers. Examples of the generated functions and the outliers are presented in Figure~\ref{fig:Fig_1}.

The functions of both the response and predictor variables in the training sample are distorted by the Gaussian random noise $\upsilon(s) \sim \mathcal{N}(0, 4)$ before fitting the model. Also, $K_{\Y} = K_{\pmb{\X}} = 10$ number of basis functions are used to smooth the noisy data. Note that in the Monte Carlo experiments, the number of basis functions $K_{\Y} = K_{\pmb{\X}} = 10$ is chosen arbitrarily to evaluate the predictive performance of the methods under the same conditions. For each scenario, MC = 250 Monte Carlo experiments are performed. In each experiment, three performance metrics; mean absolute prediction percentage error (MAPE), median absolute prediction error (MdAPE), and coefficient of determination $R^2$ are computed to evaluate the out-of-sample prediction performance of the methods:
\begin{align*}
\text{MAPE} &= \frac{1}{n_{\text{test}}} \sum_{i=1}^{n_{\text{test}}} \left\Vert \frac{\left\vert \Y_{i}(t) - \widehat{\Y}_{i}(t) \right\vert}{\left\vert \Y_{i}(t) \right\vert} \right\Vert_{\mathcal{L}_2}, \\
\text{MdAPE} &= \text{median} \left\lbrace \text{AE}_1^p, \ldots, \text{AE}_{n_{\text{test}}}^p \right\rbrace, \\
R^2 &= 1 - \frac{\sum_{i=1}^{n_{\text{test}}} \left\Vert \Y_{i}(t) - \widehat{\Y}_{i}(t) \right\Vert_{\mathcal{L}_2}^2}{\sum_{i=1}^{n_{\text{test}}} \left\Vert \Y_{i}(t) - \bar{\Y}(t) \right\Vert_{\mathcal{L}_2}^2},
\end{align*}
where $\Y_{i}(t)$ and $\widehat{\Y}_{i}(t)$ respectively denote the $i^\textsuperscript{th}$ generated and predicted functions of the response variable in the test sample and $\text{AE}_i = \vert \Y_{i}(t) - \widehat{\Y}_{i}(t) \vert$ denotes the absolute prediction error.

For further investigation of the uncertainty of predictions, the nonparametric bootstrap method is applied to construct pointwise prediction intervals for the functions of the response variable in the test sample. Let $\widehat{\epsilon}(t) = \Y(t) - \widehat{\Y}(t)$ denote the estimated model error. The following algorithm then describes the bootstrap method used in this study to construct pointwise prediction intervals for the functions of the response variable.
\begin{itemize}
\item[Step 1.] Draw a pair of bootstrap samples $\left( \Y^*(t), \pmb{\X}^*(s) \right)$ of size $n_{\text{train}}$ by sampling with replacement from the original sample $\left( \Y(t), \pmb{\X}(s) \right) $.
\item[Step 2.] Using the bootstrap sample $\left( \Y^*(t), \pmb{\X}^*(s) \right) $, obtain the IRSIMPLS approximation of $\pmb{\beta}(s,t)$, namely $\widehat{\pmb{\beta}}^{\omega*}(s,t)$ as described in Section~\ref{sec:fpls}.
\item[Step 3.] Compute the bootstrap predictions of $\Y(t)$ in the test sample as follows:
\begin{equation*}
\widehat{\Y}^*(t) = \int_S \pmb{\X}(s) \widehat{\pmb{\beta}}^{\omega*}(s,t) \textrm{d}s + \epsilon^*(t),
\end{equation*}
where $ \pmb{\X}(s)$ denotes the matrix of predictors in the test sample, and $\epsilon^*(t)$ is a random drawn from $\widehat{\epsilon}(t)$.
\item[Step 4.] Repeat Steps 1-3 $B$ times, where $B = 200$ denotes the number of bootstrap simulations, to obtain $B$ sets of bootstrap replicates of the predicted response functions $\left\lbrace \widehat{\Y}^{*,1}(t), \ldots, \widehat{\Y}^{*,B}(t) \right\rbrace$.
\end{itemize}
Denote by $Q_{\alpha}^i(t)$ the $\alpha$\textsuperscript{th} quantile of the generated $B$ sets of bootstrap replicates of the $i$\textsuperscript{th} predicted response function, where $\alpha$ denotes a level of significance. In our numerical analyses, the significance level is set to $\alpha = 0.05$ to construct 95\% prediction intervals for the functions of the response variable in the test sample. Then, $\left[ Q_{\alpha/2}^i(t), Q_{1 - \alpha/2}^i(t)\right]$ defines the $100(1-\alpha)\%$ bootstrap prediction interval for $\Y_i(t)$. For evaluating the performance of the methods, two bootstrap performance metrics; the coverage probability deviance (CPD), which is the absolute difference between the nominal and empirical coverage probabilities, and the interval score (score), are calculated as follows:
\begin{align*}
\text{CPD} &= \left|\alpha - \frac{1}{n_{\text{test}}} \sum_{i=1}^{n_{\text{test}}} \mathbb{1} \left\{  Q_{\alpha/2}^i(t) > \Y_i(t) \right\rbrace + \mathbb{1} \left\{ Q_{1 - \alpha/2}^i(t) < \Y_i(t) \right\}\right|, \\
\text{score} &= \frac{1}{n_{\text{test}}} \sum_{i=1}^{n_{\text{test}}} \bigg\Vert \left[ \left\{ Q_{1 - \alpha/2}^i(t) - Q_{\alpha/2}^i(t) \right\} \right. \\
&+ \frac{2}{\alpha} \left( Q_{\alpha/2}^i(t) - \Y_i(t) \right) \mathbb{1} \left\{ \Y_i(t) < Q_{\alpha/2}^i(t) \right\} \\
&+ \left. \frac{2}{\alpha} \left( \Y_i(t) - Q_{1 - \alpha/2}^i(t) \right) \mathbb{1} \left\{ \Y_i(t) > Q_{1 - \alpha/2}^i(t) \right\} \right] \bigg\Vert_{\mathcal{L}_2},
\end{align*}
where $\mathbb{1}\{\cdot\}$ denotes the binary indicator function.

For the aforementioned data generation process, we compare the out-of-sample predictive performance of our proposed method only with the LS and na\"ive SIMPLS-based functional PLS. The error measures for the RFPC are not calculated since it allows only one functional predictor in the model. The finite-sample performance of the proposed method will be compared with the RFPC using a different data generation process.

\begin{figure}[!htb]
  \centering
  \includegraphics[width=8.6cm]{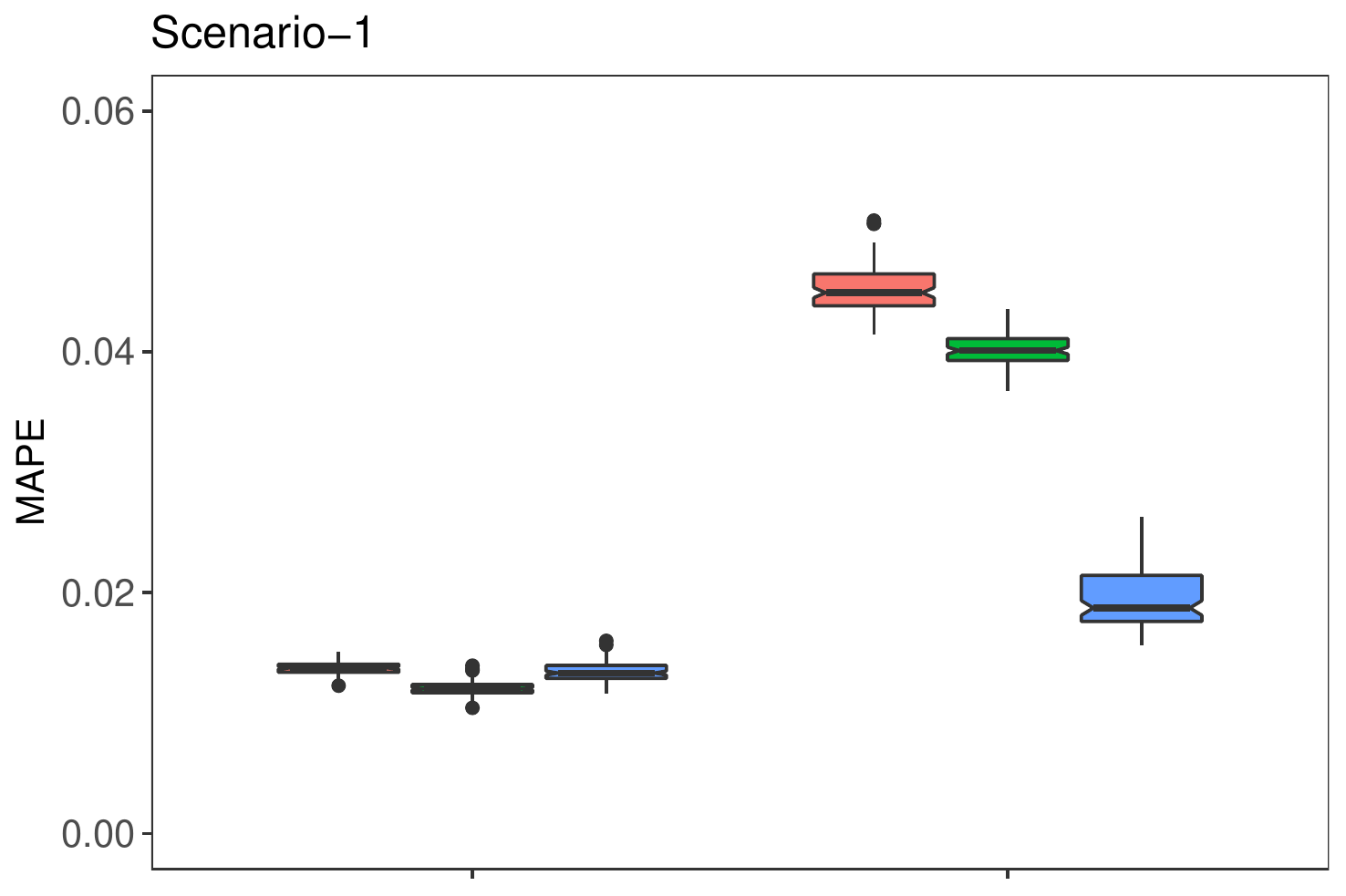}
\quad
  \includegraphics[width=8.6cm]{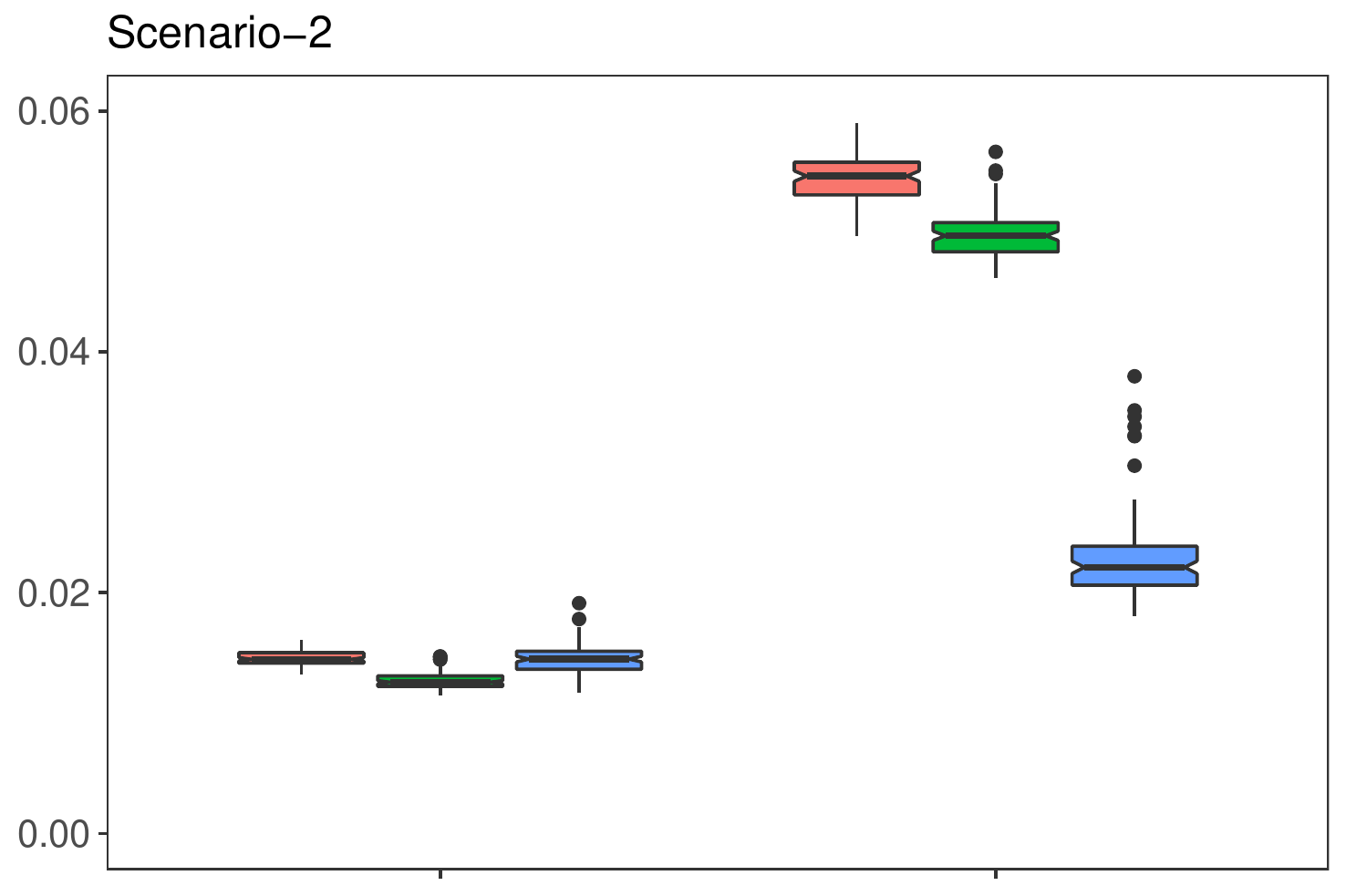}
  \\
  \includegraphics[width=8.6cm]{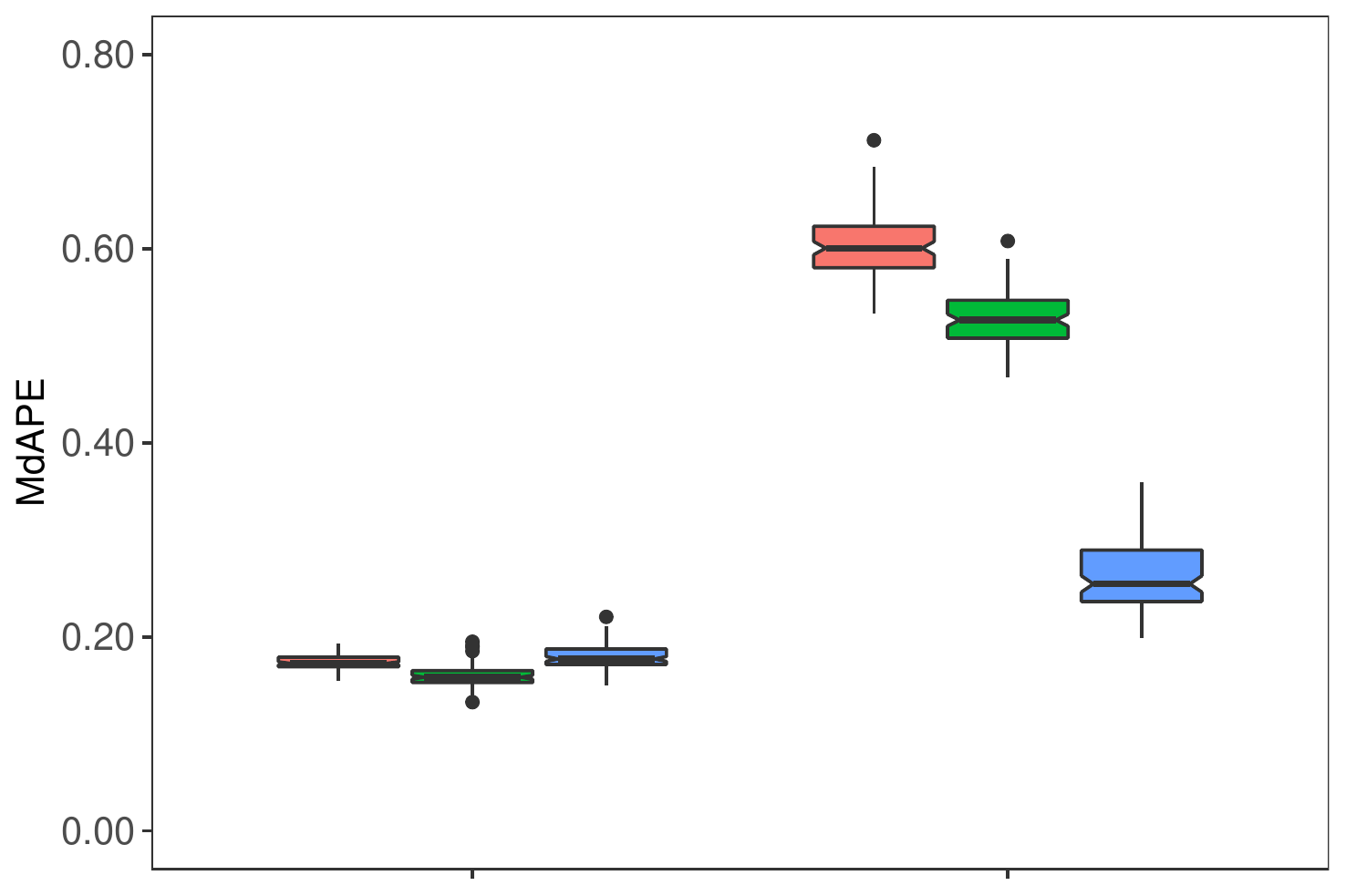}
\quad
  \includegraphics[width=8.6cm]{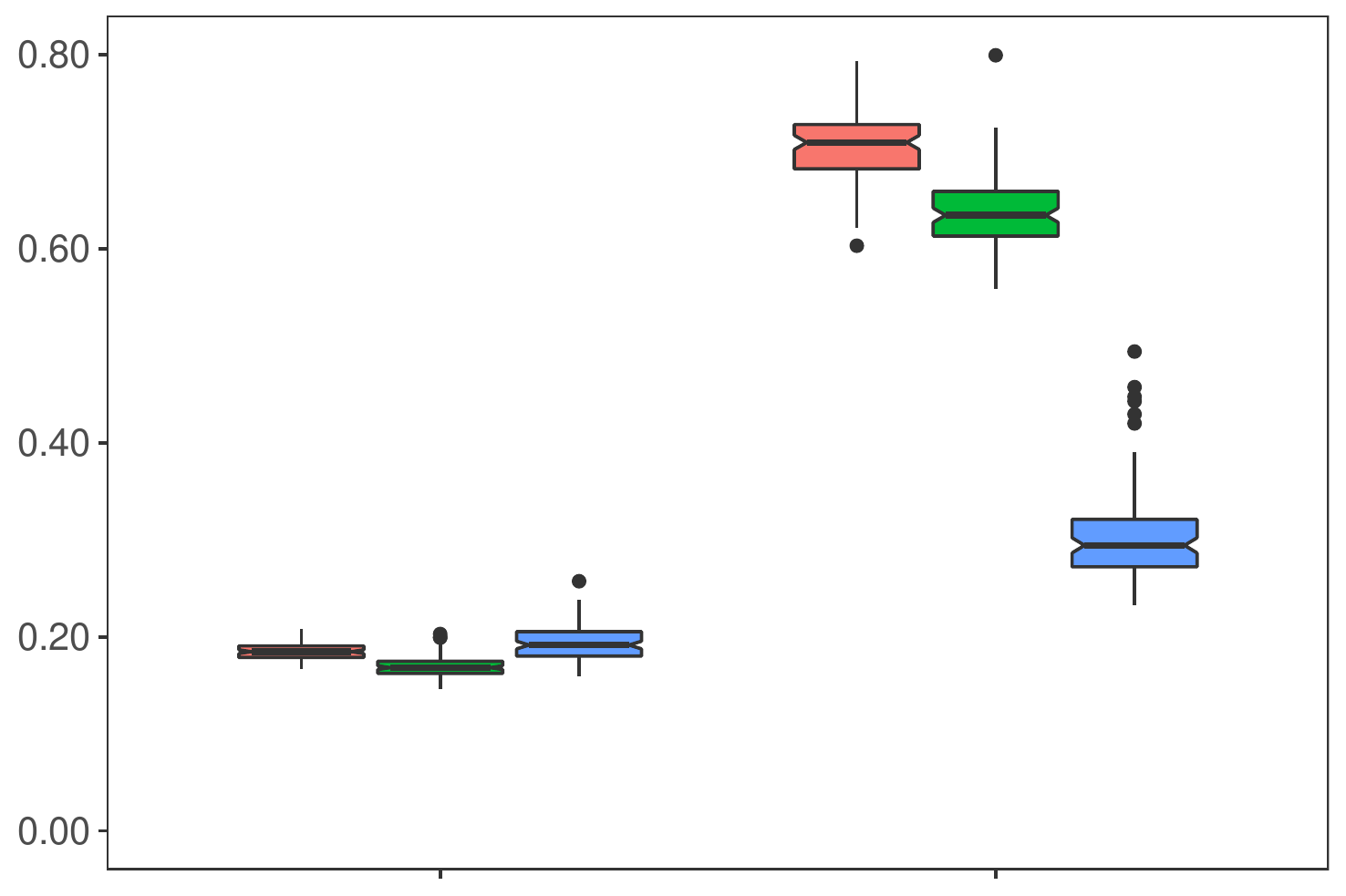}
  \\
  \includegraphics[width=8.6cm]{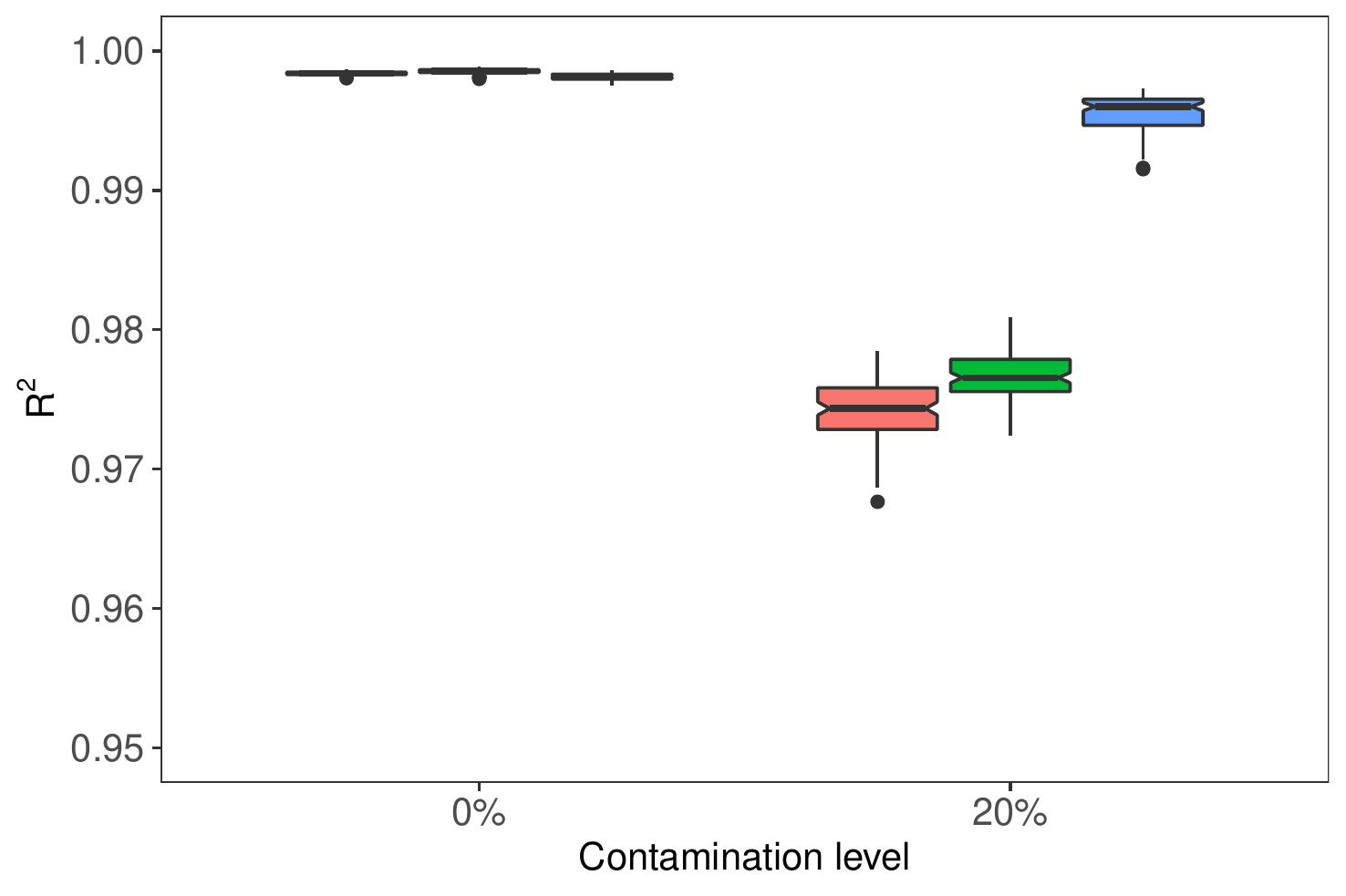}
\quad
  \includegraphics[width=8.6cm]{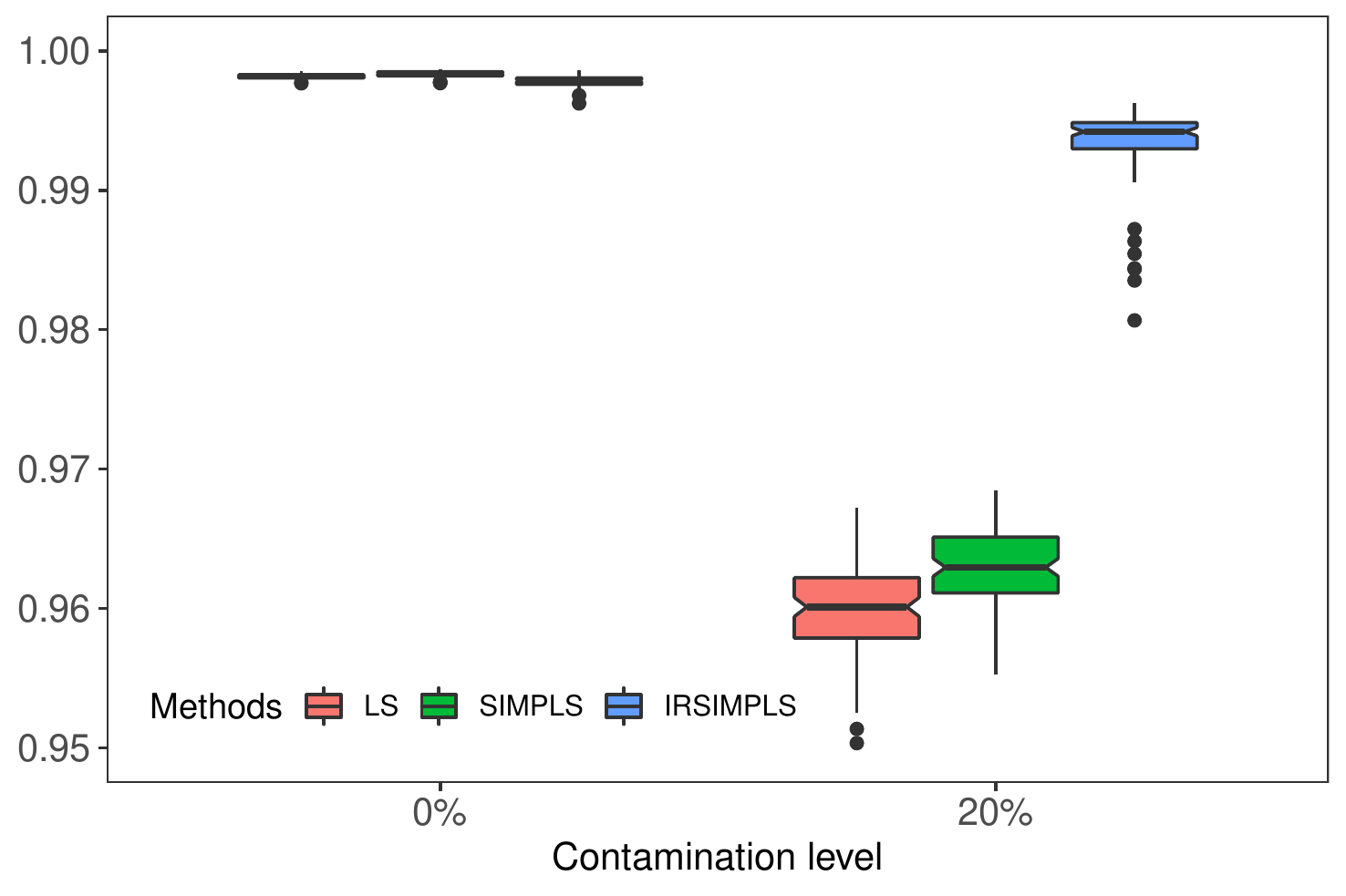}
  \caption{\small{Boxplots of the calculated MAPE (first row), MdAPE (second row), and $R^2$ (third row) values for the LS, na\"ive SIMPLS, and proposed IRSIMPLS methods under Scenario-1 (first column) and Scenario-2 (second column)}}
  \label{fig:Fig_2}
\end{figure}

\begin{figure}[!htb]
  \centering
  \includegraphics[width=8.6cm]{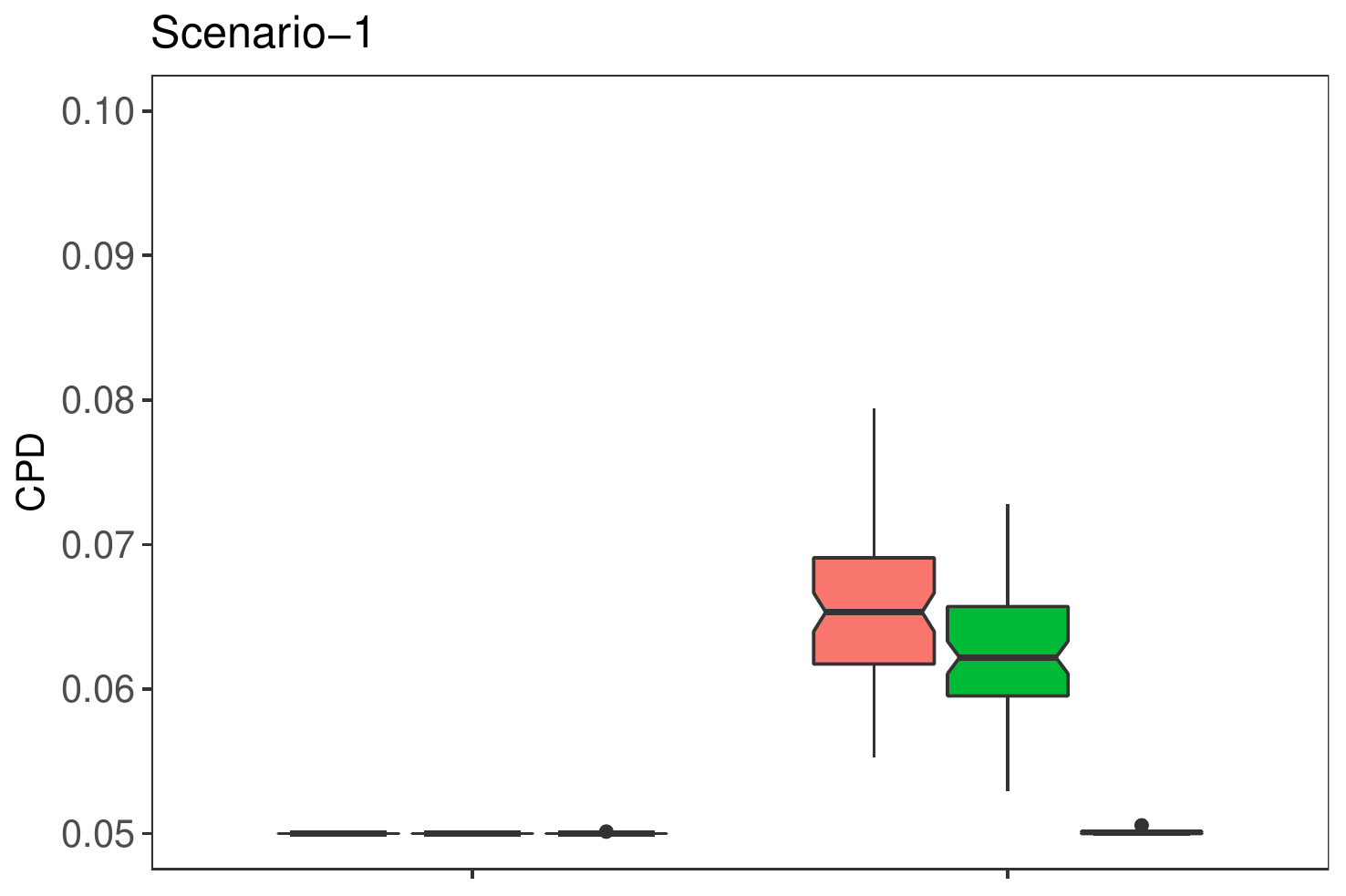}
\quad
  \includegraphics[width=8.6cm]{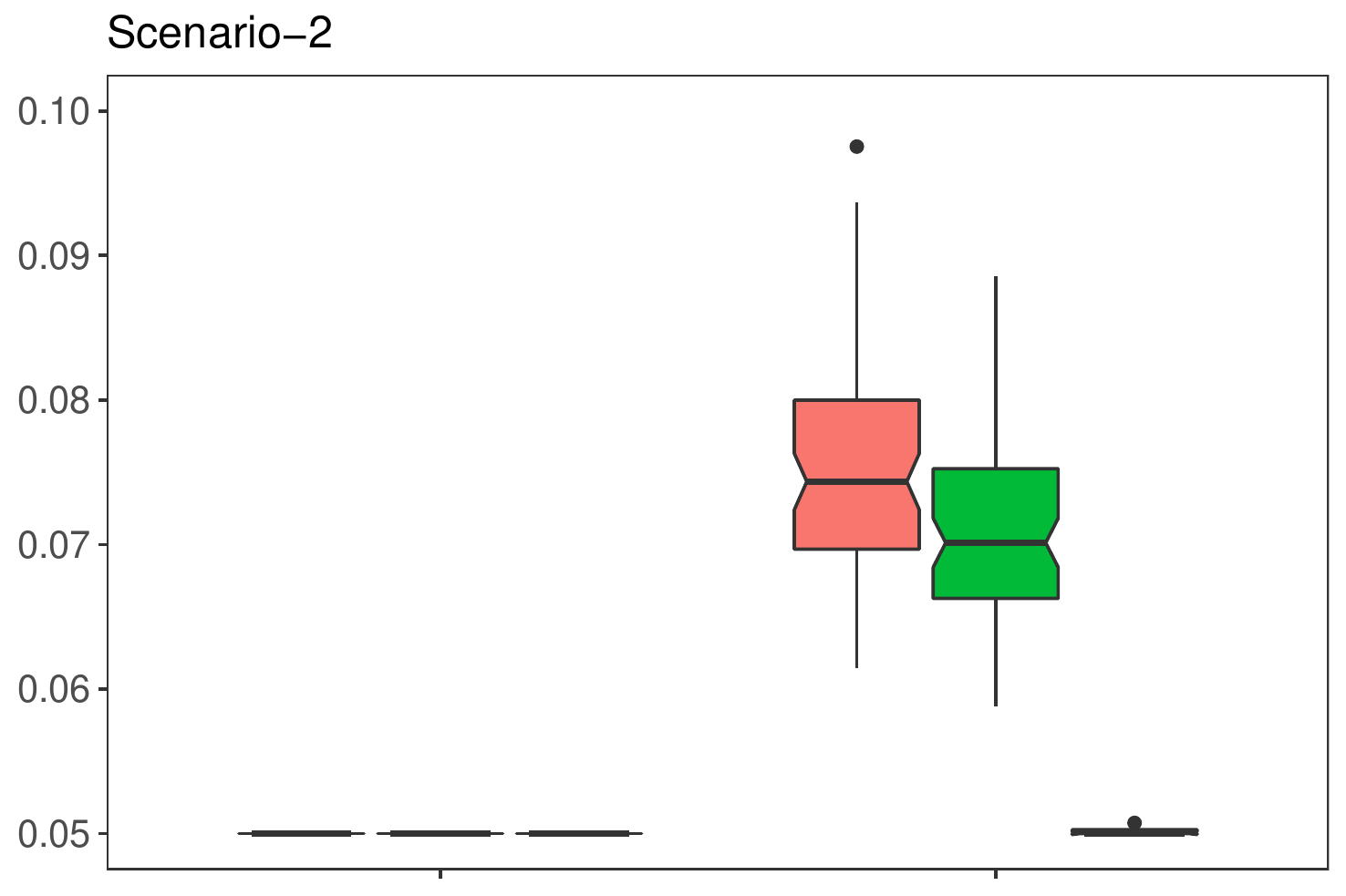}
  \\
  \includegraphics[width=8.6cm]{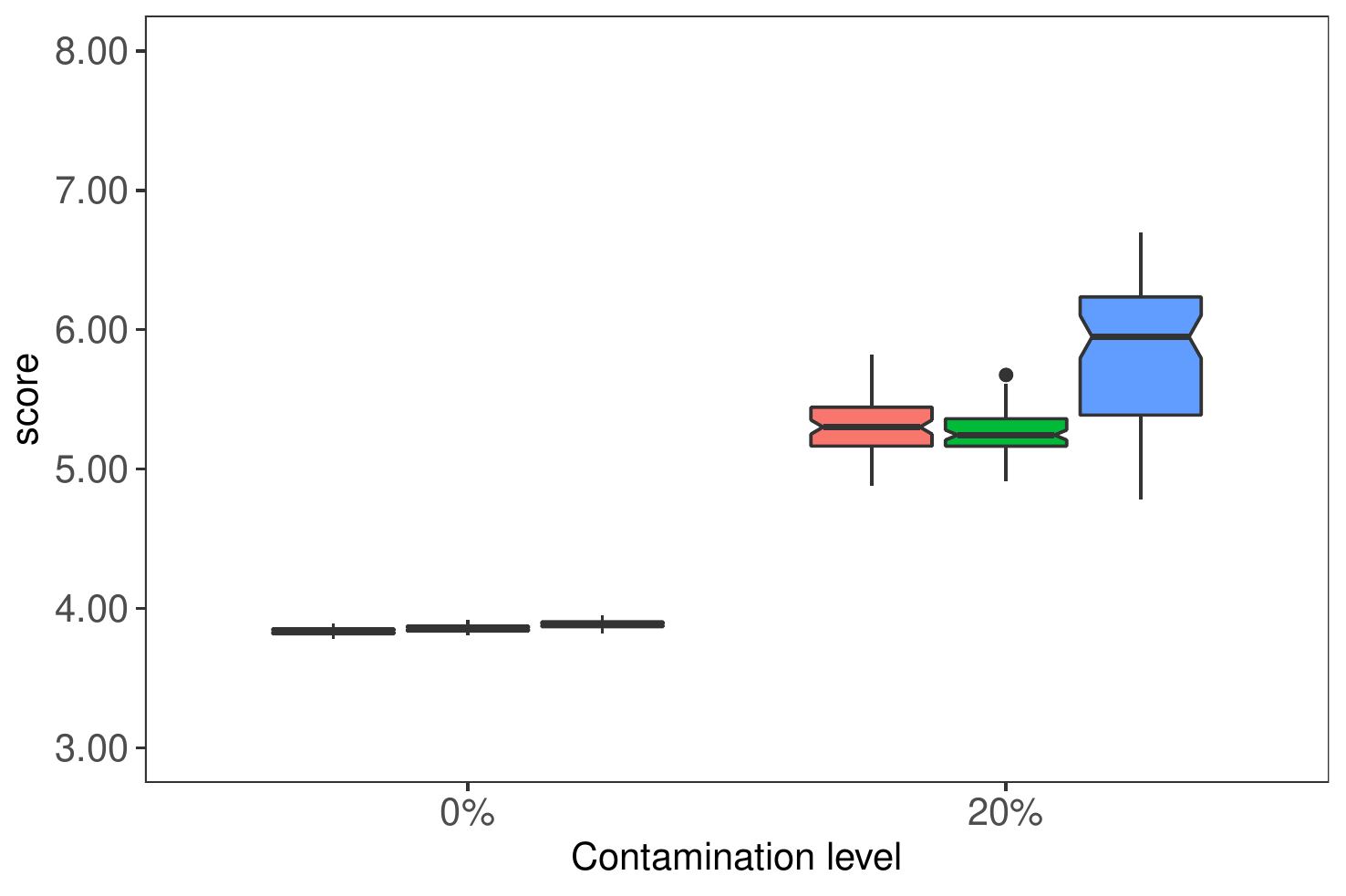}
\quad
  \includegraphics[width=8.6cm]{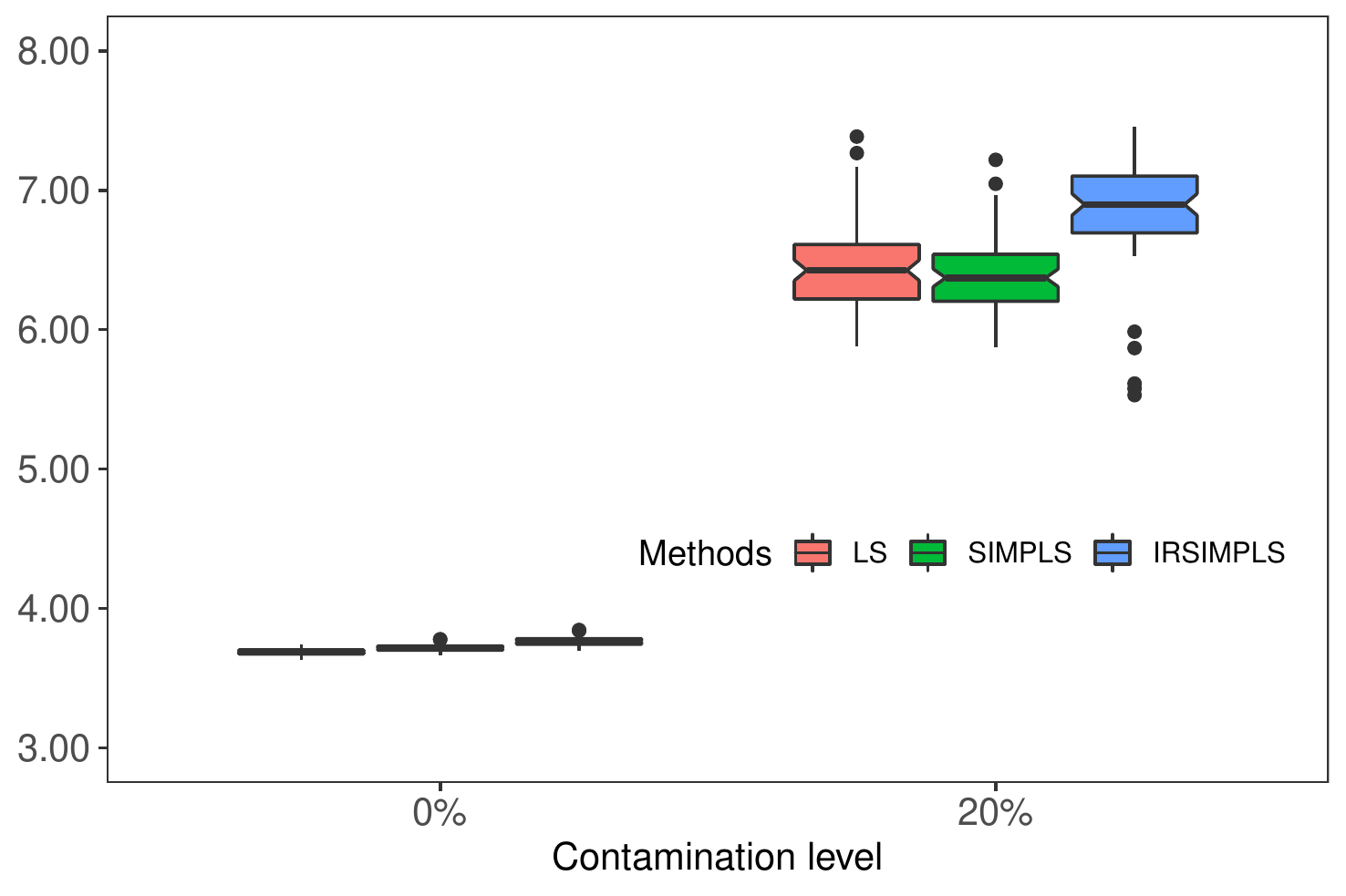}
  \caption{\small{Boxplots of the calculated CPD (first row) and score (second row) values for the LS, na\"ive SIMPLS, and proposed IRSIMPLS methods under Scenario-1 (first column) and Scenario-2 (second column)}}
  \label{fig:Fig_3}
\end{figure}

Our findings obtained from the Monte Carlo experiments are presented in Figures~\ref{fig:Fig_2} and~\ref{fig:Fig_3}. From these figures, it is obvious that all the methods produce slightly better results when the functional predictors are independent (Scenario-1) than when the functional predictors are correlated (Scenario-2). Figure~\ref{fig:Fig_2} demonstrates that when no outliers are present in the data, all the methods produce similar MAPE, MdAPE, and $R^2$ values. The traditional SIMPLS method produces slightly better performance metrics than the proposed method, but the observed differences are not significant. In addition, from Figure~\ref{fig:Fig_3}, when outliers do not contaminate the data, the proposed method produces similar bootstrap-based CPD values with slightly larger score values compared with LS and SIMPLS.

When the generated data include outliers, our proposed method produces considerably smaller MAPE, MdAPE, and larger $R^2$ values under both cases and scenarios compared with the LS and na\"ive SIMPLS. From Figure~\ref{fig:Fig_2}, the LS and na\"ive SIMPLS are affected by the outliers and produce larger error values compared with the case where outliers do not contaminate the data. On the other hand, the proposed method down-weights the effects of outliers and produces almost similar error values under both cases. Moreover, compared with LS and traditional SIMPLS, the proposed method produces significantly smaller CPD values with slightly larger score values when outliers contaminate the data (see Figure~\ref{fig:Fig_3}). In other words, compared with LS and traditional SIMPLS, much accurate prediction intervals are obtained by the proposed method when the data have outliers.

\begin{figure}[!htb]
  \centering
  \includegraphics[width=12cm]{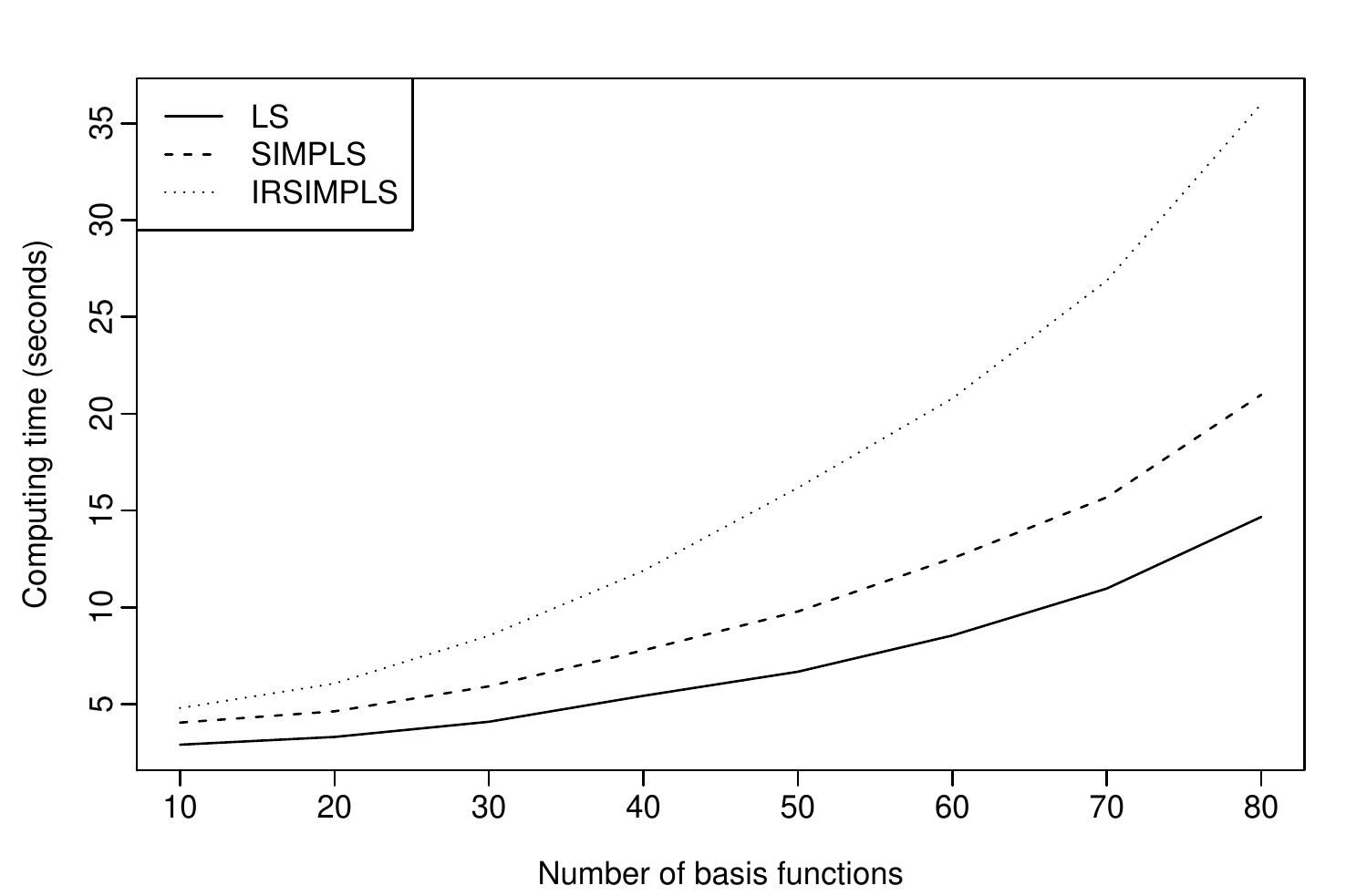}
  \caption{\small{Estimated computing times for the LS, SIMPLS, and IRSIMPLS methods}}
  \label{fig:Fig_4}
\end{figure}

Moreover, we compare the proposed method with LS and na\"ive SIMPLS in terms of their computing times. While doing so, we consider Case-1 and Scenario-1 as data generation processes and record the computing times of the methods using the \texttt{R} function \texttt{system.time} when the numbers of basis function $K_{\Y}$ and $K_{\bm{\X}}$ are increased from 10 to 80. The results are presented in Figure~\ref{fig:Fig_4}. From this figure, the proposed method requires more computing time compared with other methods. For example, the proposed method respectively requires 1.65 and 1.18 times more computing time than the LS and na\"ive SIMPLS when $K_{\Y} =K_{\bm{\X}} = 10$ while it requires 2.45 and 1.71 times more computing time when $K_{\Y} =K_{\bm{\X}} = 80$.

All in all, the numerical results produced by our Monte Carlo experiments have demonstrated that:
\begin{inparaenum}
\item[1)] the proposed IRSIMPLS-based FFLRM has a similar prediction performance to that of LS and traditional SIMPLS-based regression when no outliers are present in the data, and
\item[2)] the proposed method has improved prediction performance than the LS and traditional SIMPLS when the data are contaminated by magnitude or shape outliers. 
\end{inparaenum}
This is because the traditional approximation of $\pmb{\beta}(s,t)$ is significantly affected by outliers, leading to poor prediction performance. However, compared with LS and traditional SIMPLS, the proposed method increases the prediction performance of FFLRM by down-weighting the effects of outliers.

\begin{figure}[!htb]
  \centering
  \includegraphics[width=8.6cm]{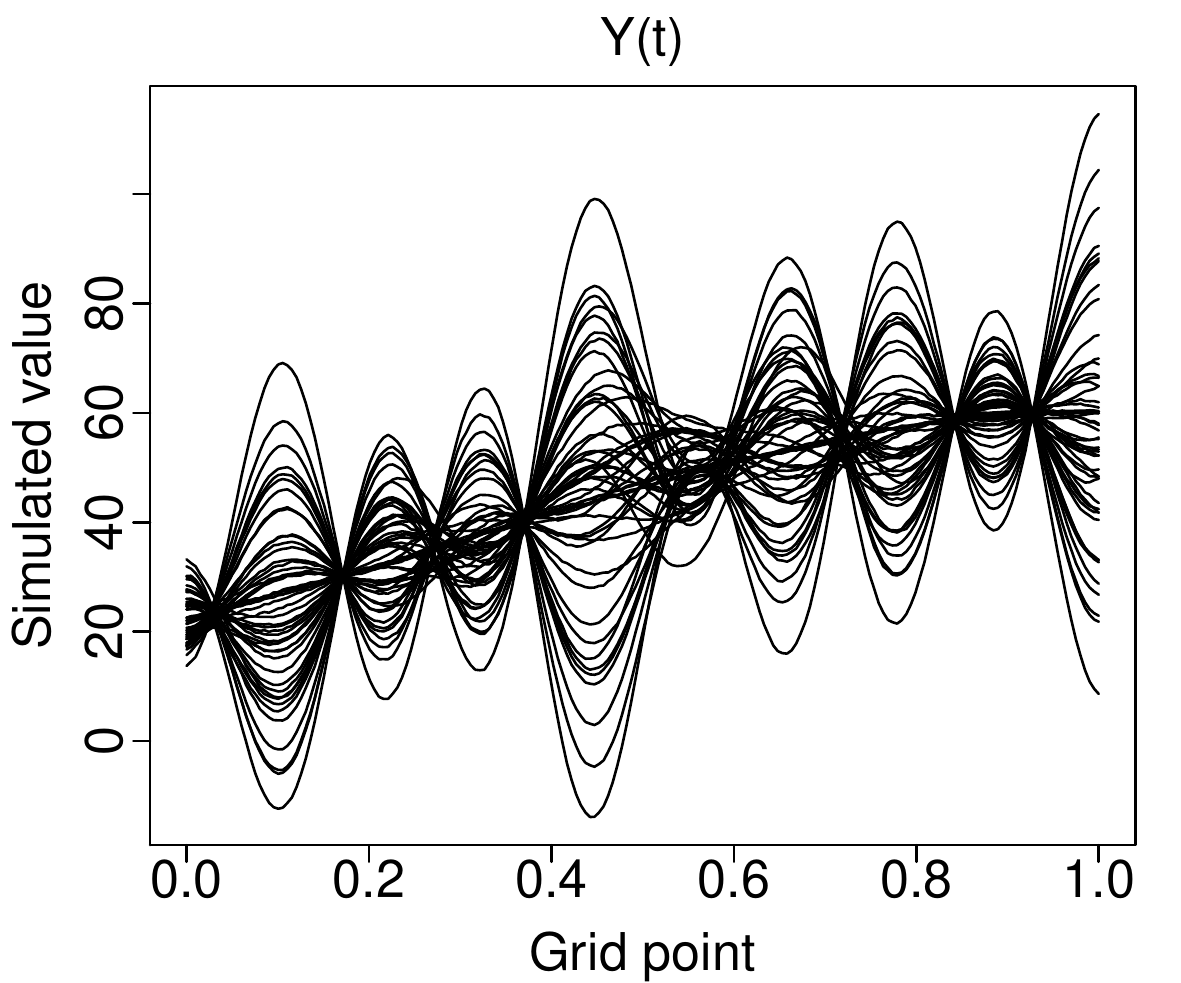}
\quad
  \includegraphics[width=8.6cm]{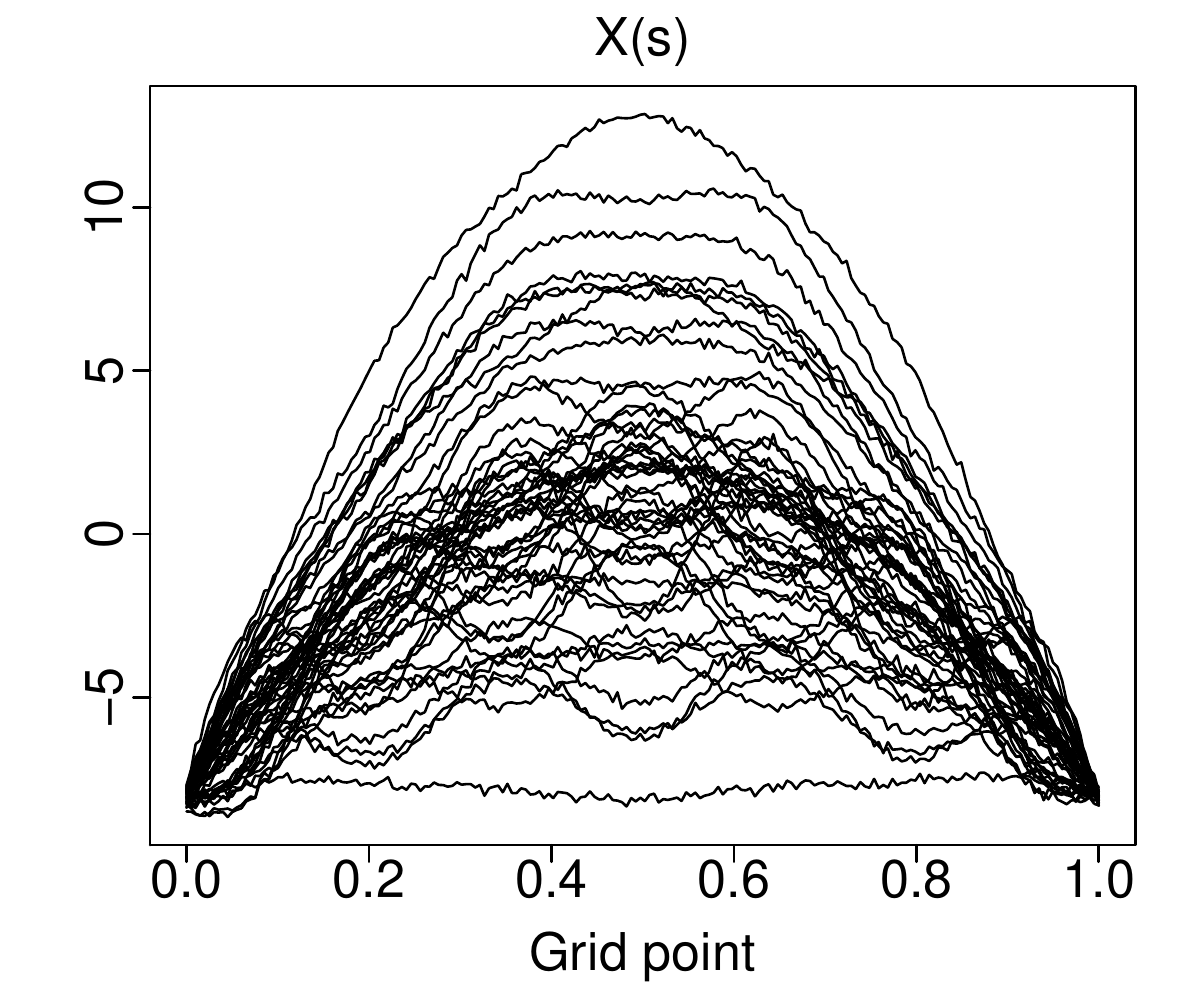}
  \caption{\small{Plots of the generated 50 functions for the functional response and predictor variables}}
  \label{fig:Fig_5}
\end{figure}

To compare the performance of our proposed method with the RFPC, we use a similar data generation process, which is a simple FFLRM, as used by \cite{Harjit}. The predictor variable $\X(s)$ is generated based on a FPC analysis with the following mean function ($\mu_{\X}(s)$) and eigenfunctions ($\iota_i^{\X}$ for $i = 1, 2, 3$):
\begin{align*}
\mu_{\X}(s) &= -10(s-0.5)^2+2, \\
\iota_1^{\X} &= \sqrt{2} \sin(\pi s),\\
\iota_2^{\X} &= \sqrt{2} \sin(7 \pi s), \\
\iota_3^{\X} &= \sqrt{2} \cos(7 \pi s),
\end{align*}
where $s \in [0,1]$ and the principal component scores are generated by sampling from the Gaussian distributions with mean-zero and variances 40, 10, and 1, respectively. The response variable $\Y(t)$, on the other hand, is generated using the following mean function ($\mu_{\Y}(t)$) and eigenfunctions ($\iota_i^{\Y}$ for $i = 1, 2, 3$):
\begin{align*}
\mu_{\Y}(t) &= 60 \exp(-(t-1)^2), \\
\iota_1^{\Y} &= \sqrt{2} \sin(12 \pi t), \\
\iota_2^{\Y} &= \sqrt{2} \sin(5 \pi t), \\
\iota_3^{\Y} &= \sqrt{2} \cos(2 \pi t),
\end{align*}
where $t \in [0,1]$. The following simple FFLRM is used to generate functional observations:
\begin{equation*}
\Y(t) = \mu_{\Y}(t) + \int_0^1 \beta(s,t) [\X(s) - \mu_{\X}(s)] \textrm{d}s + \epsilon(t),
\end{equation*} 
where $\beta(s,t) = \bm{\iota}^{\X}(s) \bm{B} \bm{\iota}^{\Y}(t)$ with $\bm{B}$ having random entries between uniform $[-3,3]$, $\epsilon(t) = \bm{q}^\top \bm{\iota}^{\Y}(t) + d$, and $\bm{q}^\top$ and $d$ are sampled from $\mathcal{N}(0, 0.1)$. We generate $n = 200$ functions for each functional variable at 200 equally spaced points in the interval $[0,1]$. The generated data are contaminated at 20\% contamination level, and two scenarios are considered to generate outliers:
\begin{description}
\item[Scenario-1] where the outliers are generated by replacing $\bm{B}$ with $\bm{B}_1 = \bm{B} + \bm{R}$ with $\bm{R} \sim \mathcal{N}(0, 0.5)$.
\item[Scenario-2] where the outliers are generated using $\beta(s,t) = \bm{\iota}^{\X}(s) \bm{B}_2[ \bm{\iota}^{\Y}(t), p(t)]$ where $\bm{B}_2$ is an $3 \times 4$ matrix $\bm{B}_2 = [\bm{B}_2, \bm{I}]$, $\bm{I}$ is sampled from $\mathcal{N}(2,1)$, and $p(t)$ is a random $B$-spline function defined on an interval of length $1/10$.
\end{description}
As noted by \cite{Harjit}, the outliers generated under Scenario-1 affect the regression function across the entire interval, whereas the outliers generated under Scenario-2 only affect a small interval of the curves. A graphical display of the generated data is presented in Figure~\ref{fig:Fig_5}.

The simple FFLRM is constructed for both scenarios, and the bivariate coefficient is estimated using the proposed method and RFPC based on $K = 15$ $B$-spline basis functions. Similar to \cite{Harjit}, the following mean squared error (MSE) is considered to compare the performance of the methods:
\begin{equation}\label{eq:mse}
\text{MSE} =  \frac{1}{(1-0.20) \times n} \sum_{i=1}^n \left\Vert (1-u_i) \widehat{\Y}_i(t) - \Y_{i}(t) \right\Vert^2_{\mathcal{L}_2},
\end{equation}
where $\widehat{\Y}_i(t)$ is the $i\textsuperscript{th}$ fitted response function and $u_i = 1$ if $i\textsuperscript{th}$ response function is outlier and $u_i = 0$ otherwise.

\begin{figure}[!htb]
  \centering
  \includegraphics[width=8.6cm]{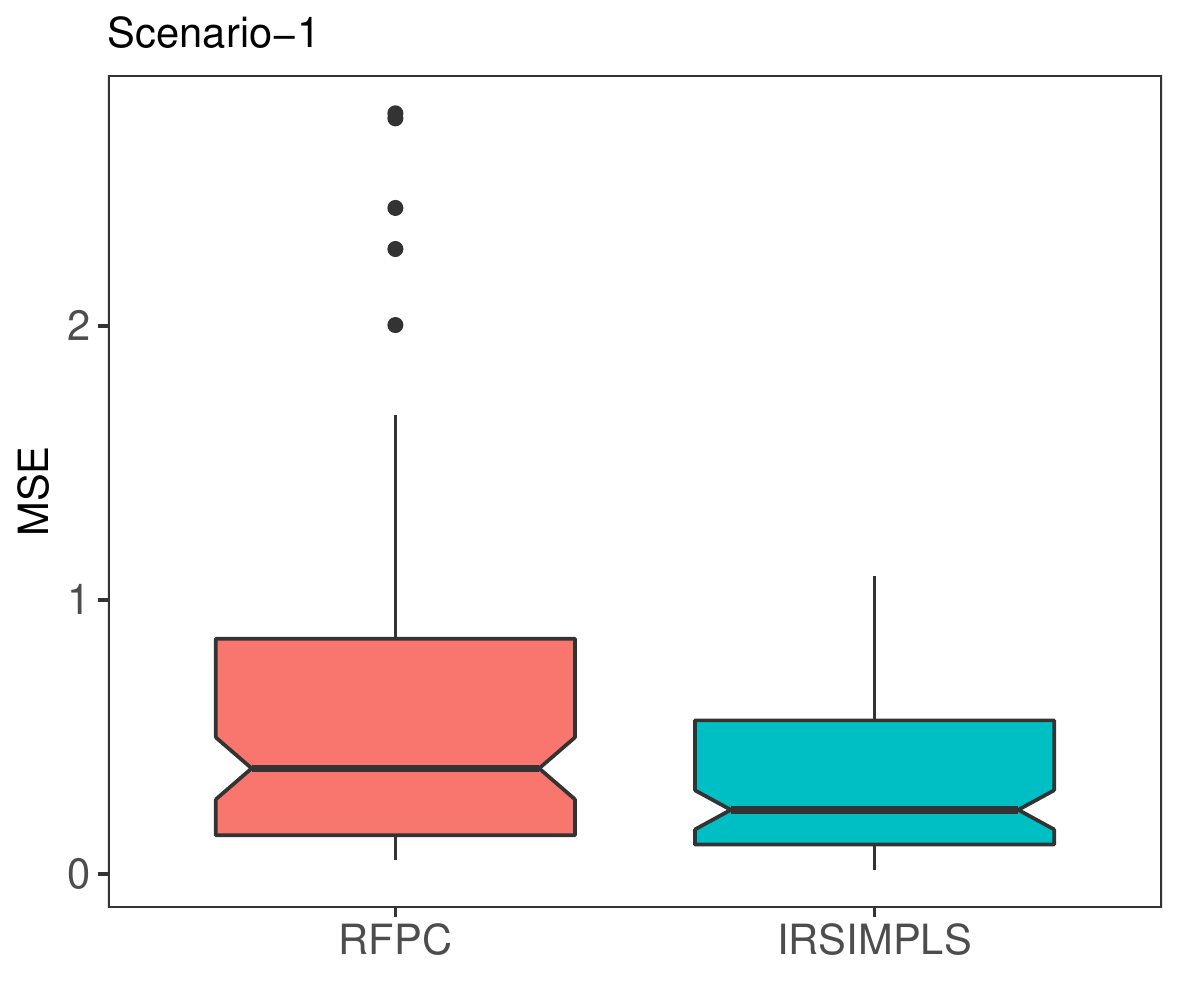}
\quad
  \includegraphics[width=8.6cm]{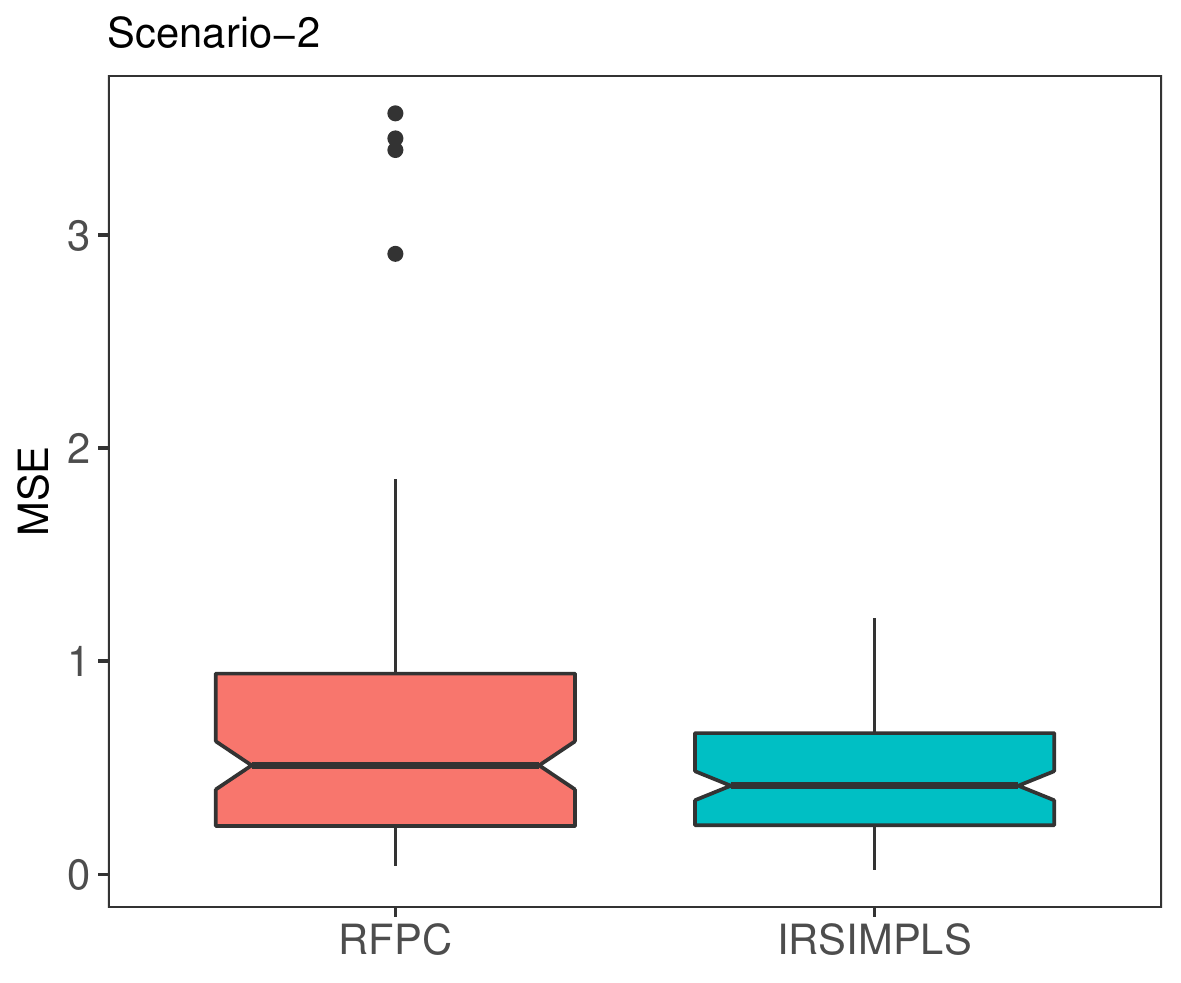}
  \caption{\small{Boxplots of the calculated MSE values for the RFPC and proposed IRSIMPLS methods, when the data are generated under Scenario-1 (left panel) and Scenario-2 (right panel)}}
  \label{fig:Fig_6}
\end{figure}

The MSE values obtained from 100 replications are presented in Figure~\ref{fig:Fig_6}. From this figure, our proposed method produces smaller MSE values compared with the RFPC of \cite{Harjit}.

\section{Empirical data example: Oman weather data}\label{sec:eda}

\begin{figure}[!htb]
  \centering
  \includegraphics[width=5.6cm]{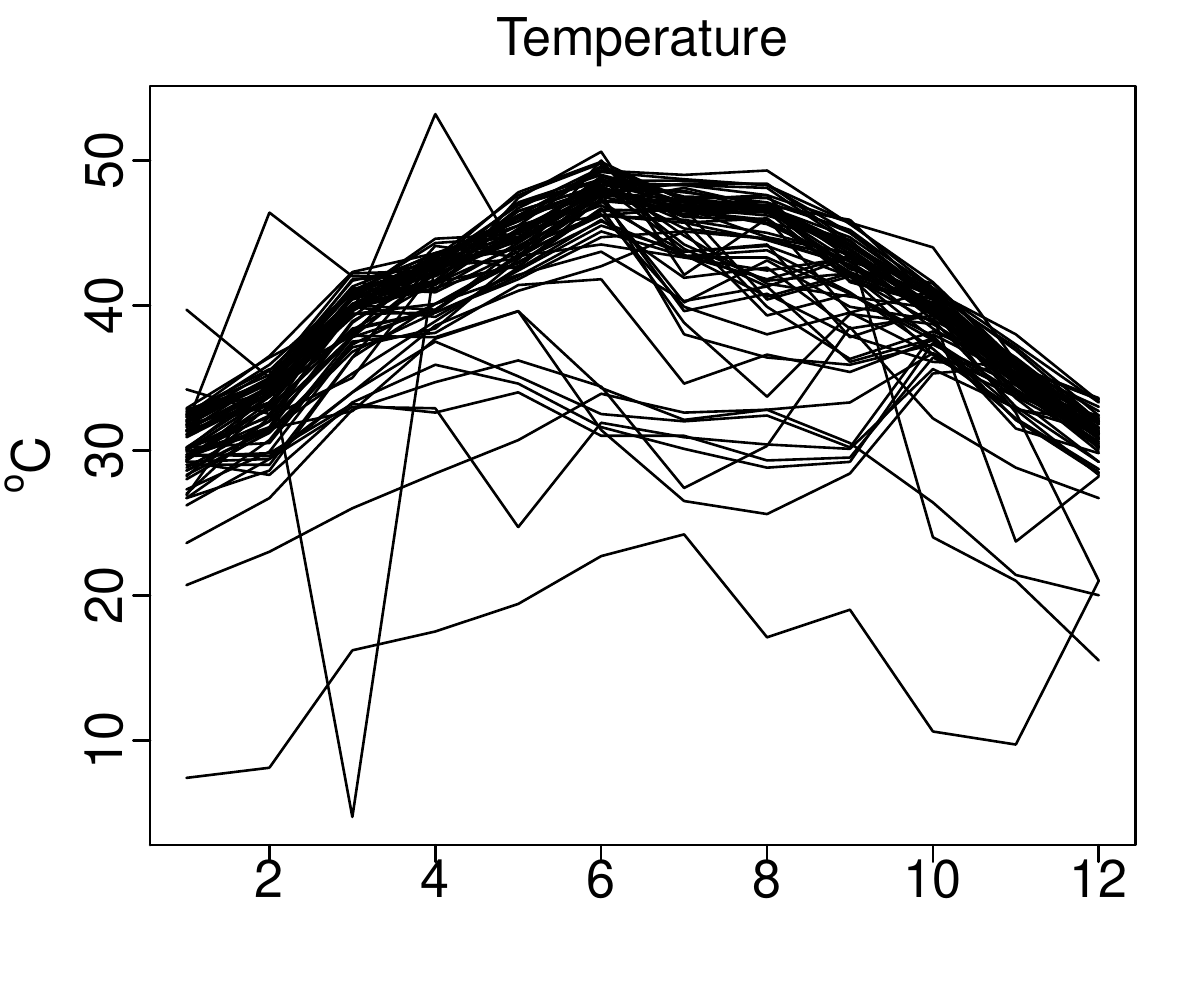}
  \includegraphics[width=5.6cm]{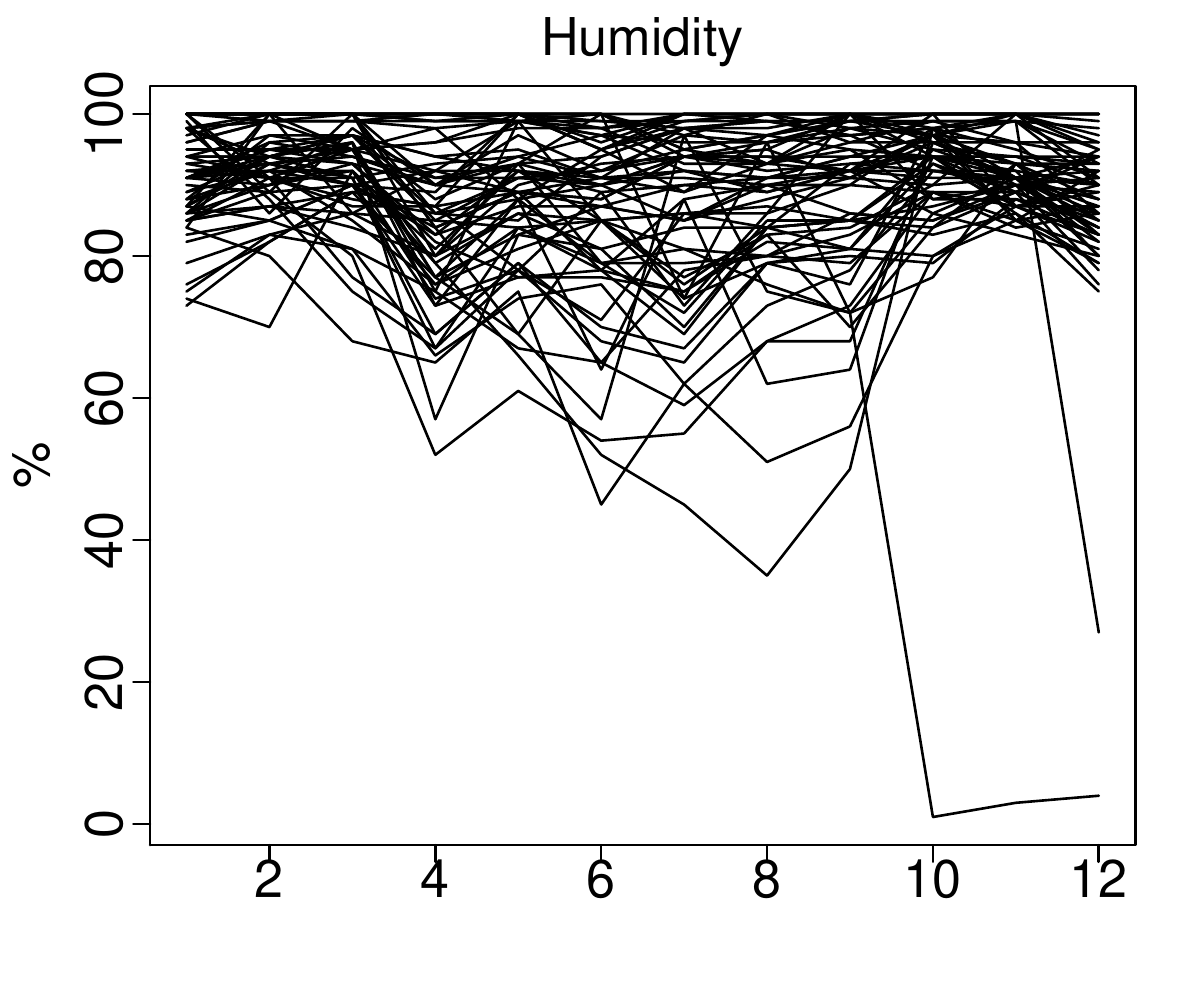}
  \includegraphics[width=5.6cm]{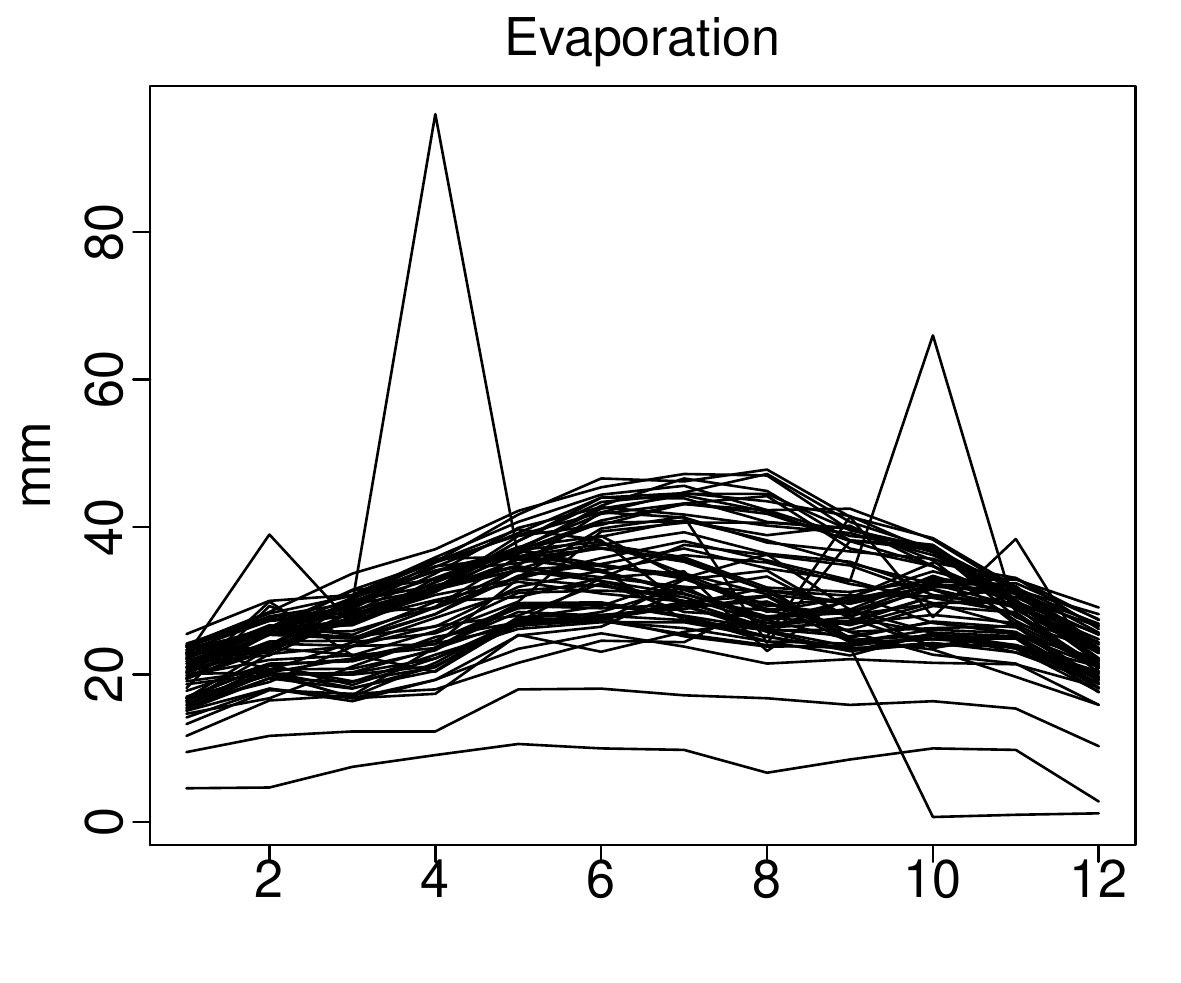}
  \\
  \includegraphics[width=5.6cm]{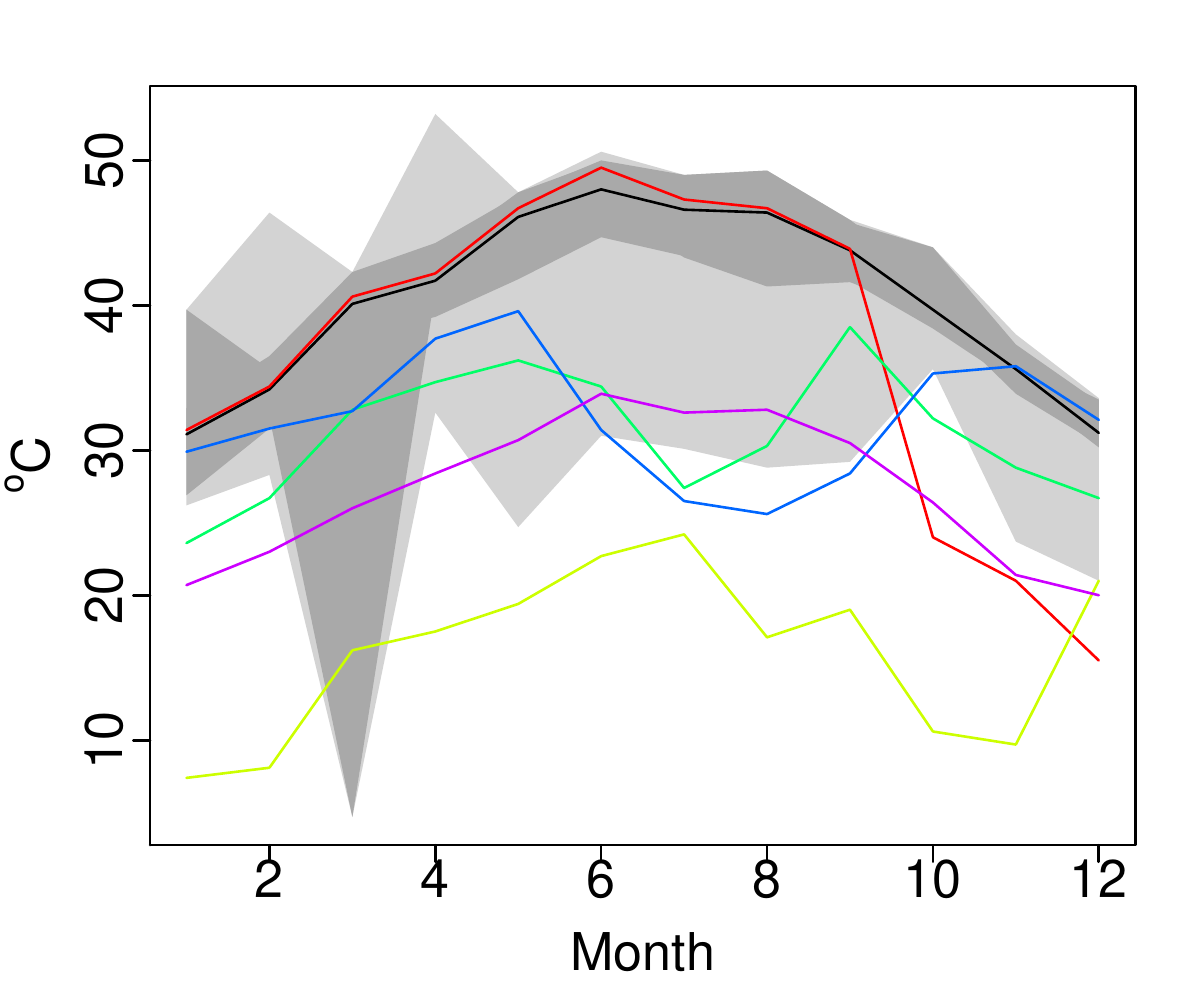}
  \includegraphics[width=5.6cm]{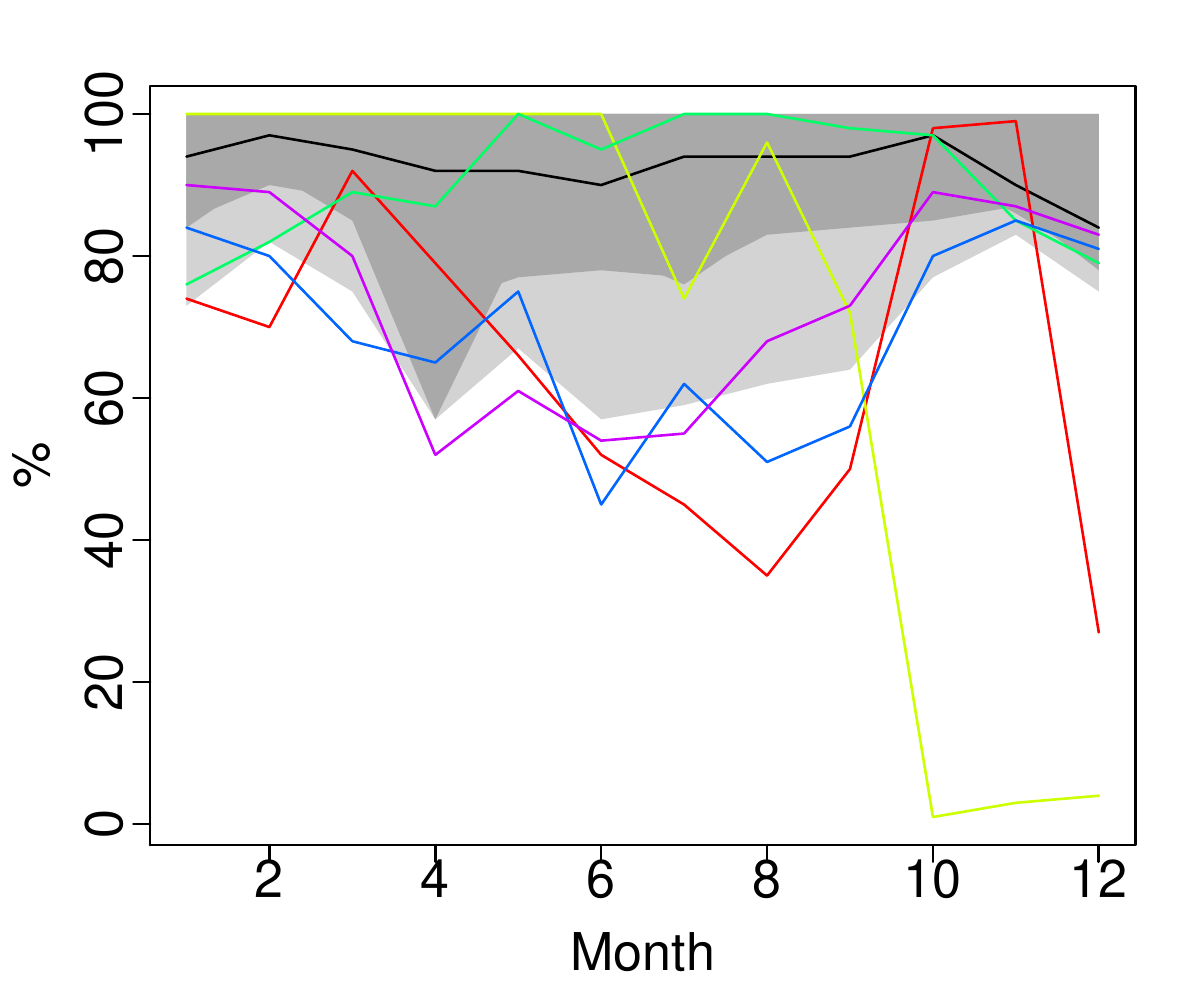}
  \includegraphics[width=5.6cm]{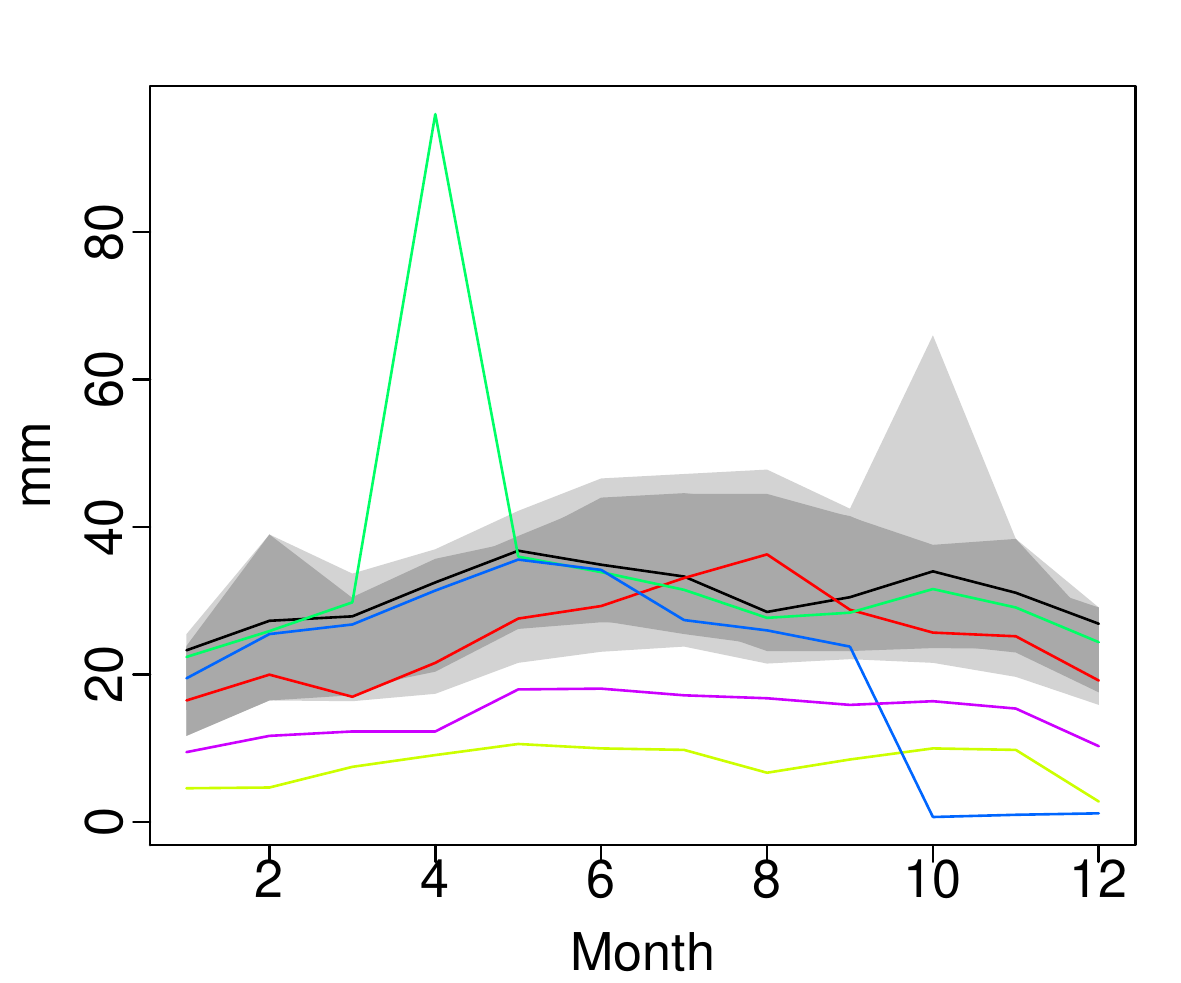}
  \caption{\small{Graphical display of the temperature, humidity, evaporation variables (first row), and their HDR boxplots (second row) for the Oman weather data. The observations are the functions of months and $1 \leq s,t \leq 12$}}
  \label{fig:Fig_7}
\end{figure}

The finite-sample performance of the proposed method is also evaluated using the Oman weather empirical dataset (dataset is available from the National Center for Statistics \& Information: \url{https://data.gov.om}). The dataset contains three variables: maximum monthly evaporation (mm), humidity (\%), and temperature ($^\circ$C), which were collected from 66 weather stations across Oman from January 2018 to December 2018. The observations are considered as the functions of months ($1 \leq s,t \leq 12$). Figure~\ref{fig:Fig_7} presents the graphical display of the variables, as well as the functional highest density region (HDR) boxplots \citep{hdr} for these variables. In the functional HDR boxplots, the inner and outer regions are presented with dark and light gray colors for each variable, respectively. The inner region corresponds to the region bounded by all the curves corresponding to the points inside the 50\% bivariate HDR. In contrast, the outer region corresponds to the region bounded by all the curves corresponding to the points within the outer bivariate HDR \citep{hdr}. Functions that are outside of these regions are called outliers. From this figure, all the variables have clear outlying curves.

For this dataset, we consider predicting maximum monthly evaporation using the maximum temperature and humidity variables. The following procedure is repeated 100 times to evaluate the predictive performance of the methods. The dataset is randomly divided into the training and test samples with sizes 40 and 26, respectively. The model is constructed based on the curves in the training sample using $K_{\Y} = K_{\pmb{\X}} = 6$ number of basis functions, which are determined based on the generalized cross-validation procedure of \cite{craven}, to predict 26 curves in the test sample. For each replication, the MAPE, MdAPE, and $R^2$, as well as the bootstrap-based score and CPD values, are calculated. 

\begin{figure}[!htb]
  \centering
  \includegraphics[width=5cm]{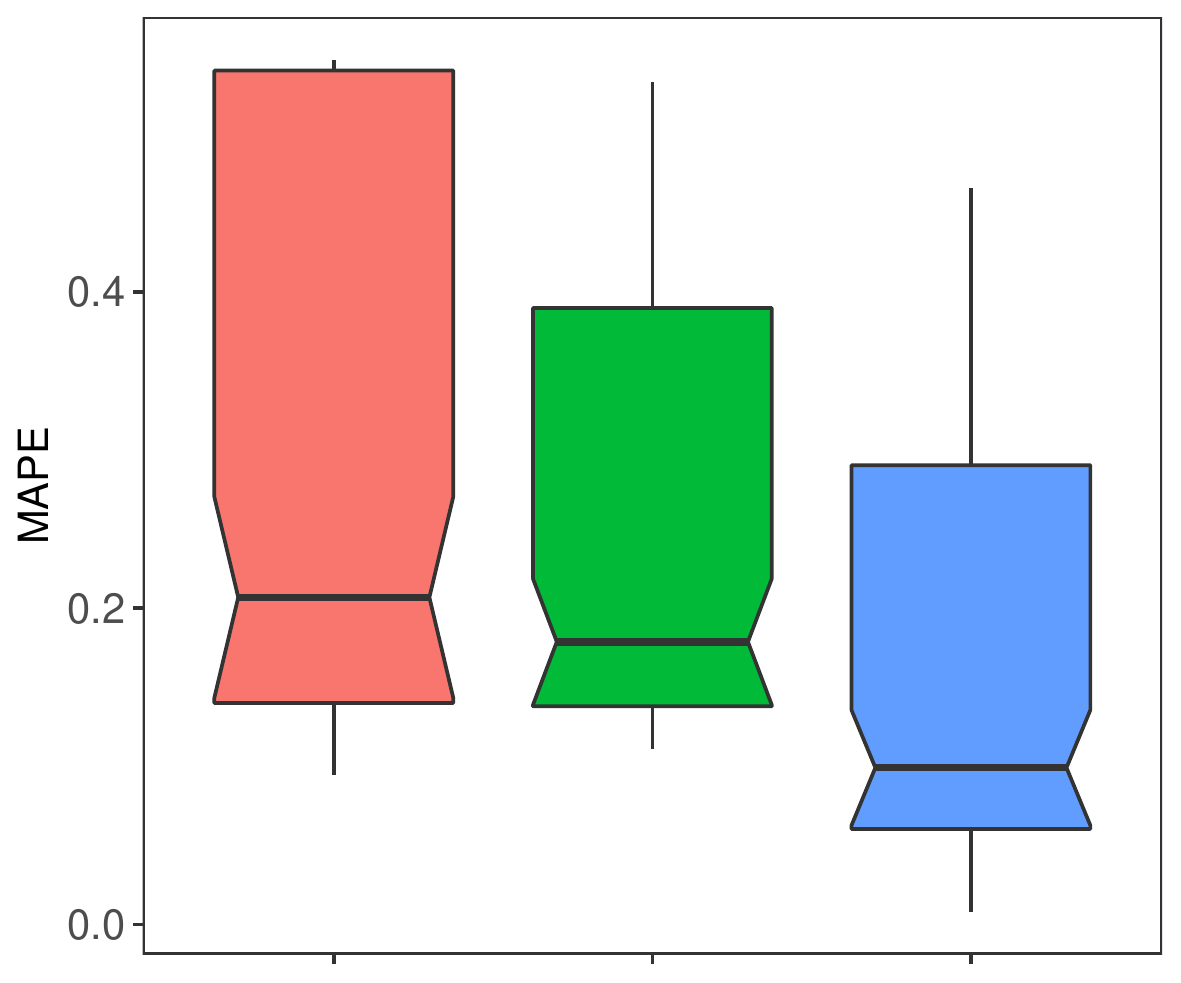}
  \includegraphics[width=5cm]{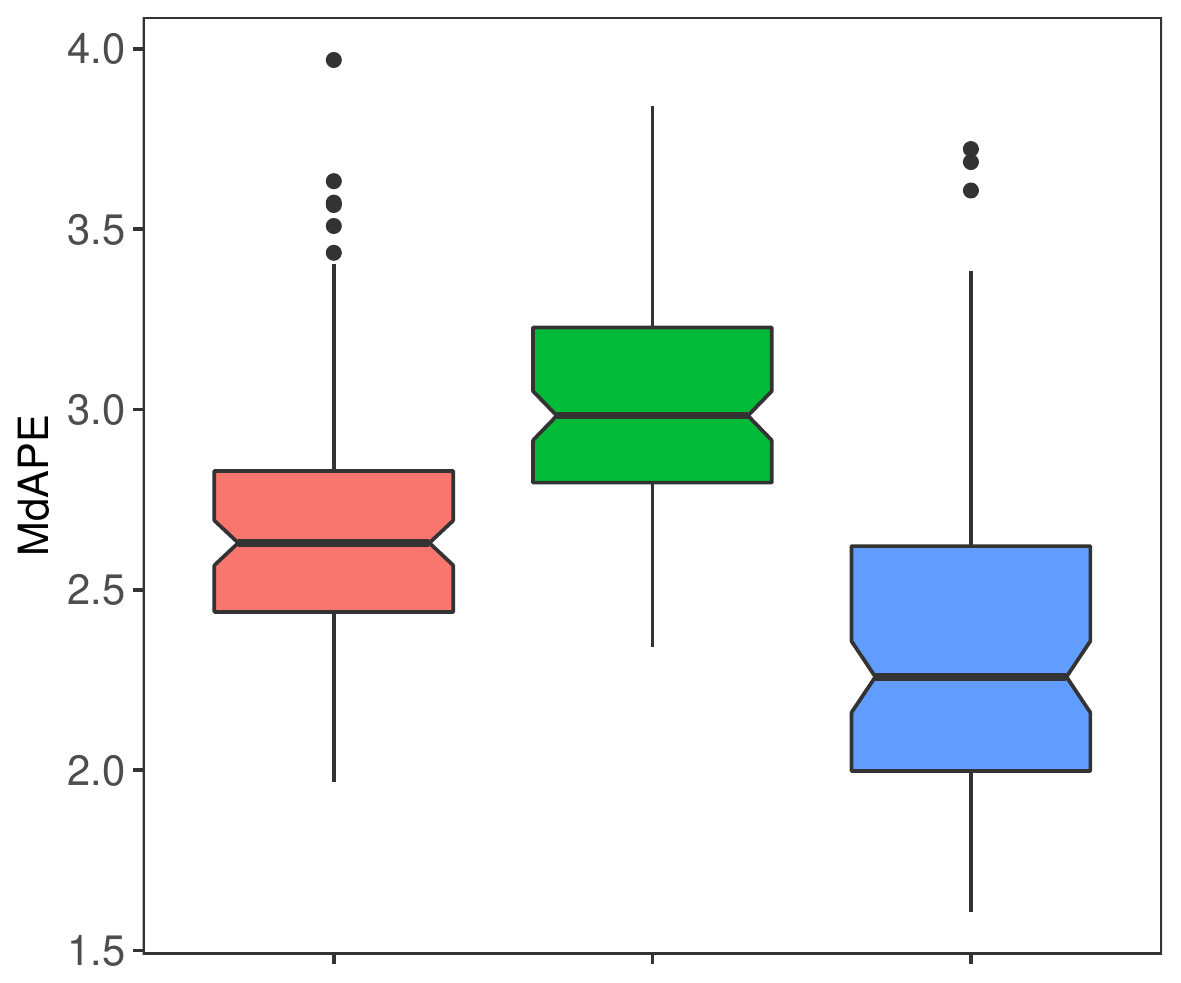}
  \includegraphics[width=5cm]{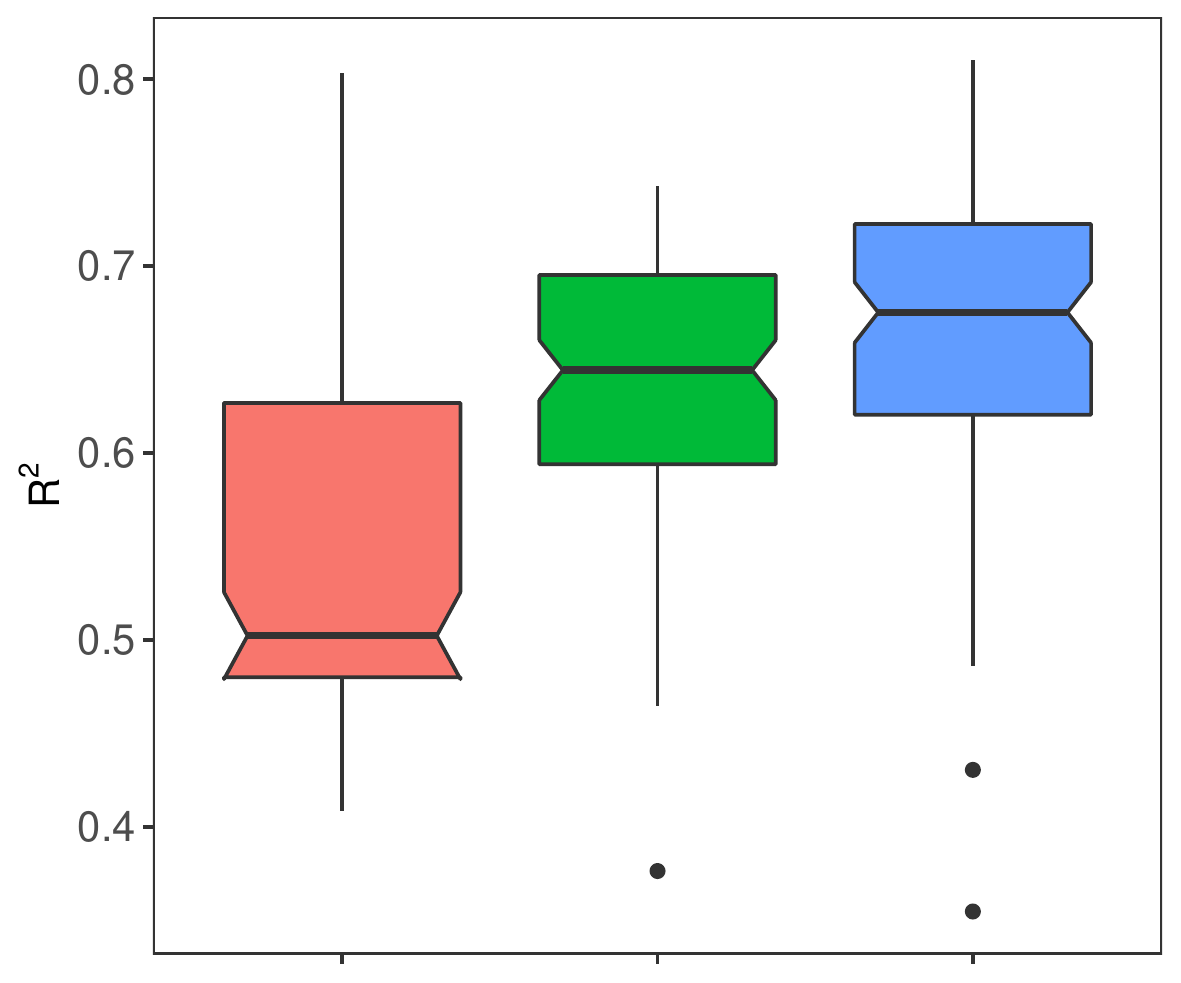}
  \\
  \includegraphics[width=5cm]{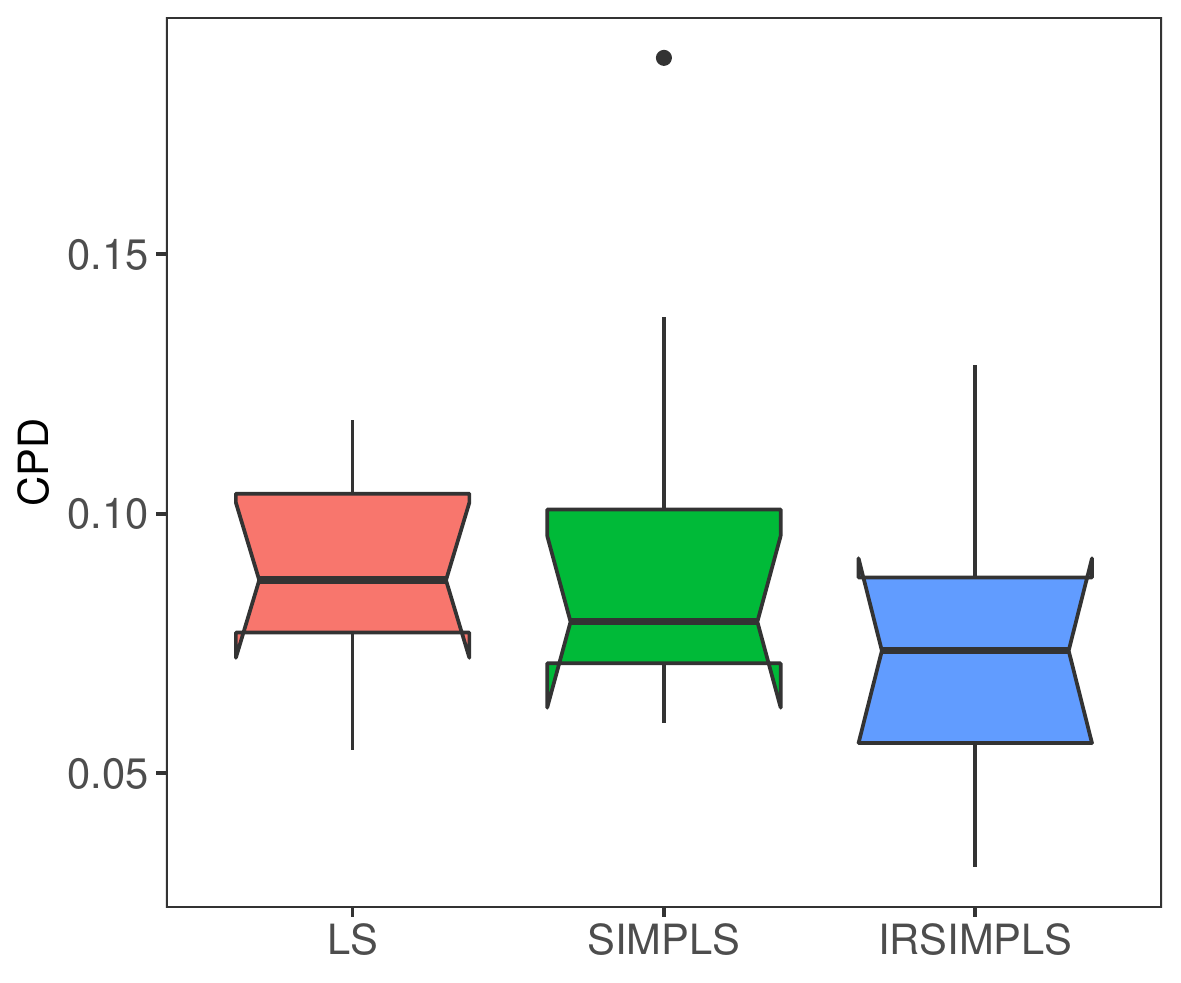}
  \includegraphics[width=5cm]{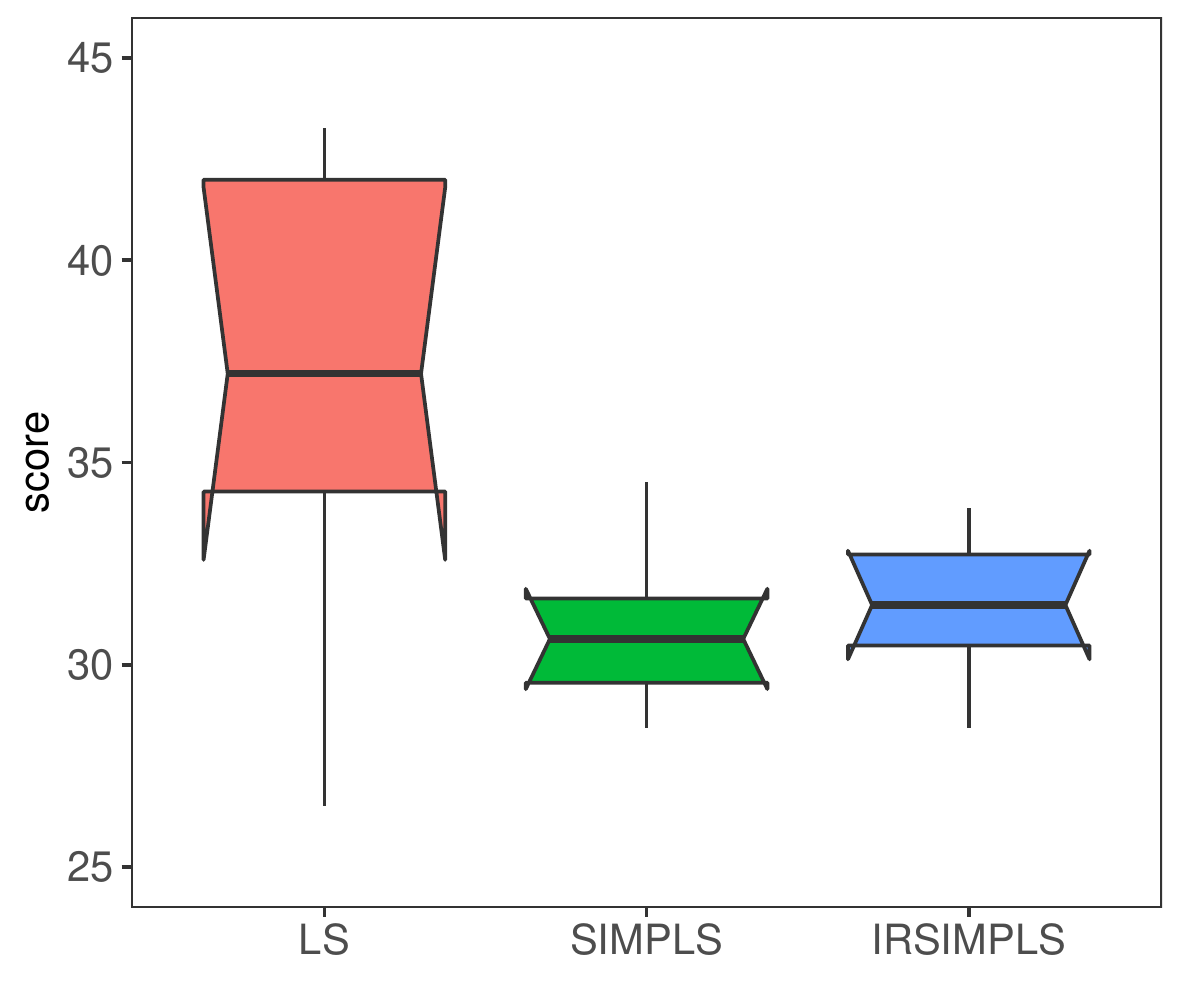}
  \caption{\small{Computed MAPE, $R^2$, bootstrap-based CPD, and score values for the LS, traditional SIMPLS, and proposed IRSIMPLS methods for the Oman weather data}}
  \label{fig:Fig_8}
\end{figure}

Our results are presented in Figure~\ref{fig:Fig_8}, which indicates that the proposed IRSIMPLS-based method produces better prediction performance compared with the LS and na\"ive SIMPLS. The proposed method has smaller MAPE and MdAPE and larger $R^2$ values than LS and classical SIMPLS. In addition, the proposed method produces smaller bootstrap-based CPD values than the LS and classical SIMPLS. However, it produces slightly larger score values than classical SIMPLS, as presented in Figure~\ref{fig:Fig_8}.

To compare the performance of our method with the RFPC, we consider two simple FFLRMs. In the first model, we consider temperature as the predictor and evaporation as the response variable, whereas humidity is the predictor variable in the second model. We construct models based on the entire data and calculate the MSE values using the observed and fitted curves as in~\eqref{eq:mse}. The computed MSE values for the first model are 21.75 and 17.87 for the RFPC and IRSIMPLS, respectively. For the second model, the computed MSE values are observed as 20.43 and 21.98. Our proposed method can produce competitive performance compared with the RFPC for the Oman weather data.

\section{Conclusions, limitations, and future research} \label{sec:conc}

FFLRMs have become an important analytical tool to explore the relationships between complex and high-dimensional functional variables. Accordingly, several FFLRMs have been proposed. The traditional estimation techniques, including PLS regression, have been extended to $\mathcal{L}_2$ Hilbert space to estimate the coefficient function of the considered regression models. However, these techniques may be significantly affected by the presence of outliers. In this case, traditional methods produce biased estimates for the regression parameter, leading to poor prediction results.

In this study, we propose a robust PLS method based on the IRSIMPLS approach of \cite{AA17} to obtain an outlier-resistant estimate for the coefficient function of the FFLRM. Its predictive performance is evaluated using various Monte Carlo experiments and an empirical data example. The results are compared with the LS, na\"ive SIMPLS, and the RFPC method of \cite{Harjit}. In addition, we applied a nonparametric bootstrap to obtain pointwise prediction intervals for the functions of the response variable. Our findings demonstrate that the proposed method produces similar results to LS and na\"ive SIMPLS when the data do not include outliers. On the other hand, our proposed method produces better prediction performance than the LS and traditional SIMPLS when the data contain outliers. The numerical results obtained from the Monte Carlo experiments performed in this study have shown that the proposed method produces more accurate bootstrap prediction intervals with wider prediction interval lengths than the traditional methods in the presence of outliers. Another interesting result produced by our Monte Carlo experiments is that, similar to the traditional methods, the predictive performance of the proposed method is affected by the correlation level between the predictors. Compared to the case where the functional predictors are correlated, our proposed method produces better predictive performance when the predictors are independent. However, it still produces improved performance over competitors when the predictors are correlated, and outliers contaminate the data. Contrary to the superiority of the proposed method over the traditional methods, it requires more computing time than the LS and na\"ive SIMPLS methods since the proposed method uses an iterative weighted likelihood algorithm. The results of our Monte Carlo experiments and empirical data analysis have also shown that the proposed IRSIMPLS method produces improved estimation and prediction accuracy than RFPC.

There are several directions that our proposed method can be further extended. For example:
\begin{inparaenum}
\item[1)] When a relatively large number of functional predictors are used in the FFLRM, not all of them may significantly affect the dependent variable. In such a case, the IRSIMPLS-based FPLS method may produce poor predictive performance. Thus, similar to \cite{LuoQi} and \cite{BS21}, a variable selection procedure along with the IRSIMPLS can be used to improve the predictive performance of the proposed method.
\item[2)] Our numerical analyses consider only the $B$-spline basis function expansion to project the infinite-dimensional random functions into the finite-dimensional space. The predictive performance of the proposed method may be dependent on the type of basis functions under consideration. As a future study, the finite-sample performance of our proposed method can be explored using other available basis function expansion methods, such as Fourier, wavelet, radial, and Bernstein polynomial bases.
\item[3)] In this study, we only consider the main effects of the predictors on the response variable. However, recent studies \citep[see, e.g.,][]{LuoQi, matsui2020, BS21} have demonstrated that the FFLRM with quadratic and interaction effects of the functional predictors perform better than the standard FFLERM. On the other hand, the outliers in the quadratic and interaction effects of the predictors may be more erroneous than the effects of outliers in the main effects. The proposed method can also be extended to the FFLRM with quadratic and interaction effects to estimate this model robustly.
\item[4)] Our numerical analyses consider a standard bootstrap method to construct a pointwise prediction interval for the response variable. While our Monte Carlo experiments show that the proposed method produces accurate prediction intervals in the presence of outliers, it produces wider prediction intervals than traditional methods. A robust bootstrap method such as the one proposed by \cite{BeyaztasJaS} can be used along with the proposed method to construct accurate prediction intervals with narrower prediction interval lengths.
\item[5)] One of the recent studies in the functional data analysis literature is about the misaligned observations \citep{Olsen2019}. When misaligned functional observations are present in the data, they may be considered as shape outliers, and such observations may lead to poor modeling results. The proposed method can also be extended to the analysis of misaligned functional observations to reduce the effects of such observations on the analysis results.
\end{inparaenum}

\section*{Acknowledgements}

This work was supported by The Scientific and Technological Research Council of Turkey (TUBITAK) (grant no: 120F270). We thank Dr. Aylin Alin for providing some R code.

\newpage
\bibliographystyle{agsm} 
\bibliography{robust_pls}       

\end{document}